\documentclass[%
 aip,
 jcp,%
 amsmath,amssymb,
reprint,%
]{revtex4-1}

\usepackage{dcolumn}
\usepackage{color,soul}

\usepackage{graphicx}
\usepackage{dcolumn}
\usepackage{bm}
\usepackage[normalem]{ulem}

\begin{document}


\title{A comparative study of different machine learning methods for dissipative quantum dynamics}

\author{Luis E. Herrera Rodr\'iguez}
\affiliation{Department of Physics and Astronomy, University of Delaware, Newark, DE 19716, United States}
\affiliation{Departmento de F\'isica, Universidad Nacional de Colombia, Bogot\'a D. C., Colombia}
\author{Arif Ullah}%
\affiliation{State Key Laboratory of Physical Chemistry of Solid Surfaces, Fujian Provincial Key Laboratory of Theoretical and Computational Chemistry, Department of Chemistry, and College of Chemistry and Chemical Engineering, Xiamen University, Xiamen 361005, China}
\author{Kennet J. Rueda Espinosa}
\affiliation{Department of Physics and Astronomy, University of Delaware, Newark, DE 19716, United States}
\affiliation{Departmento de F\'isica, Universidad Nacional de Colombia, Bogot\'a D. C., Colombia}
\author{Pavlo O. Dral}
\email{dral@xmu.edu.cn}
\affiliation{State Key Laboratory of Physical Chemistry of Solid Surfaces, Fujian Provincial Key Laboratory of Theoretical and Computational Chemistry, Department of Chemistry, and College of Chemistry and Chemical Engineering, Xiamen University, Xiamen 361005, China}
\author{Alexei A. Kananenka}
\email{akanane@udel.edu}
\affiliation{Department of Physics and Astronomy, University of Delaware, Newark, DE 19716, United States}

\date{\today}

\begin{abstract}
It has been recently shown that supervised machine learning (ML) algorithms can accurately and efficiently 
predict the long-time populations dynamics of dissipative quantum systems given only short-time
population dynamics. In the present article we benchmaked 22 ML models on their ability to predict long-time dynamics of
a two-level quantum system linearly coupled to harmonic bath. The models include uni- and bidirectional
recurrent, convolutional,
and fully-connected feed-forward artificial
neural networks (ANNs) and kernel ridge regression (KRR) with linear and most commonly used nonlinear kernels.
Our results suggest that KRR with nonlinear kernels
can serve as inexpensive yet accurate way to simulate long-time dynamics in cases where the constant
length of input trajectories is appropriate. Convolutional Gated Recurrent Unit model is found to be
the most efficient ANN model. 
\end{abstract}

\maketitle

\section{\label{sec:intro}Introduction}
Simulation of the dynamics of quantum dissipative
systems~\cite{weiss12,breuer02,leggett87} is one of the most challenging problems in physics and
chemistry. Quantum dissipation arises from the coupling of
a quantum system to a thermal bath which consists of an infinite
number of degrees of freedom. This results in a time irreversible dynamics of the system. 

A multitude of numerically exact methods has been developed for quantum dynamics simulations 
 including the hierarchical equations of motion (HEOM),~\cite{tanimura89a,tanimura20} 
multi-configurational time-dependent Hartree (MCTDH),~\cite{meyer90,wang03} quasi-adiabatic propagator path integral
(QUAPI),~\cite{makarov94} time evolving density matrix using orthogonal polynomials algorithm (TEDOPA),~\cite{prior10}
time-dependent density matrix renormalization group (TD-DMRG),~\cite{ren18} time-dependent
Davydov ansatz,~\cite{luo10} diagrammatic quantum Monte Carlo,~\cite{cohen11,cohen15}
tensor-train split-operator Fourier transform (TT-SOFT),~\cite{greene17} and the stochastic equation of motion approach.~\cite{yan2016,hsieh18,hsieh18a,han2020,ullah2020} 
In practice, however, none of these methods can be used to simulate long-time quantum dynamics of
quantum systems coupled to realistic baths containing large number of, generally, anharmonic degrees of freedom (DOF). 

Methods based on Nakajima--Zwanzig generalized quantum master equation (GQME)~\cite{nakajima58,zwanzig60b,shi03,kelly13,mulvihill21,mulvihill21a,brian21}
including transfer tensor method (TTM),~\cite{cerrillo14,kananenka16,buser17,gelzinis17,chen20a} 
allow to obtain long-time dynamics of the reduced density matrix of a quantum system
at a significantly lower computational cost compared to numerically exact methods, 
provided the GQME memory kernels
are available. 
However, in general, it is difficult to numerically
calculate the exact memory kernel for GQME. TTM further reduces 
the computational cost compared to the direct solution of GQME but it still requires an
input set of dynamical maps 
to be generated by a numerically accurate method rendering such approaches 
 infeasible for systems with many DOFs. 
Machine learning (ML) offers an alternative route to accurate, yet 
greatly accelerated quantum dynamics calculations with minimum input information 
required.~\cite{rodriguez21,ullah21,ullah2022predicting,ullah2022one} 

In ML, predicting the future time-evolution of quantum mechanical 
observables based on past values can be formulated as
time series forecasting problem.
A time series is a set of data points recorded
in consecutive time intervals. Forecasting is a challenging part of time series data analysis.
Traditional approaches to forecasting of time series utilize moving averages. One such approach, 
Autoregressive Integrated Moving Average (ARIMA) is a linear regression-based method
which, together with its variations (e.g., ARIMAX and SARIMA),~\cite{box2015time} 
have become the standard tools for modeling various time series problems.~\cite{ariyo14,Khashei11} 
These approaches perform reasonably
well for short-term forecasting, but the performance of these methods
deteriorates severely for long-term predictions.
Additionally, such methods require assumptions about the underlying data
that have to be incorporated into the model. 

Machine learning is artificial intelligence-based 
data analysis technique that is data-driven as opposed
to traditional model-driven approaches.
Perhaps, the most widely known ML tool is a feed-forward neural network (FFNN)
which is an artificial neural network (ANN) wherein connections between the nodes do not form a 
cycle.~\cite{schmidhuber15,goodfellow2016deep}
In this kind of ANN the information moves only in one direction from the input nodes, 
through the hidden nodes (if any) and to the output nodes without using any feedback from
the past input data.  As a result, the output of any layer does
not affect the training process performed on the same layer. 
These types of ANNs are, in general, not
 able to handle sequential inputs and all their inputs
(and outputs) must be independent of each others.

In contrast, recurrent neural networks (RNNs),~\cite{sherstinsky20,hochreiter97,goodfellow2016deep} 
in general, are designed with built-in
gates that function to store the previous inputs and
leverage sequential information. RNNs 
utilizes a loop in the network to preserve some information
and thus functions like a memory. 
This gives such feedback-based models the ability to 
learn from past data. RNNs are known to outperform
ARIMA-based models in the long-time forecasting
problems.~\cite{ye12,siaminamini18,siaminamini19}

RNNs have been successfully applied to time series 
classification~\cite{malhotra17,fawaz19,kashiparekh19} and forecasting
across many domains including financial data predictions,~\cite{moghar20,fisher18,kim19,kumar21}
speech~\cite{graves2012supervised,graves05,graves12,graves13,hannun14,xiong17} and handwriting
recognition,~\cite{graves07,graves09} natural language processing,~\cite{tang16,yin17} machine translation,~\cite{sutskever14,bahdanau16,shido19}
healthcare,~\cite{choi17,lynn19,gupta21,ma17}
traffic speed prediction,~\cite{vanlint05,zheng17,cui19} music,~\cite{eck02,boulanger12} video,~\cite{srivastava15}
meteorology,~\cite{habi19} molecular drug discovery,~\cite{gupta18,segler18} 
demand forecasting,~\cite{wei19,abbasimehr20} gaming,~\cite{zhang20}
remote sensing~\cite{haobo16,ienco17} computer code generation~\cite{bhoopchand16} and others.

There are several variations of RNN-based models differing in their capabilities to remember input data.
The vanilla RNN,~\cite{goodfellow2016deep,pineda87} 
 long short-term memory (LSTM),~\cite{hochreiter97,graves05,gers00,gers99,goodfellow2016deep}
and gated recurrent unit (GRU)~\cite{cho14,goodfellow2016deep} are the most commonly used
types of RNNs.
The vanilla RNN is the first model of recurrent ANNs that was introduced.~\cite{pineda87} 
Its ability to learn
long-term dependencies is limited due to vanishing and exploding gradient problems.
LSTM-based models are extensions of RNNs with significantly improved
performance on long-term predictions.  
LSTM uses a set of gates to learn what information worth to remember.
GRU is another gated variant of RNN model developed by Chung \textit{et al.}~\cite{chung14}
who showed that GRU outperforms the
LSTM in some tasks. Josefowicz \textit{et al.}~\cite{jozefowicz15} also reported the 
superior performance of the GRU-based models, but noticed that the performance
of LSTM can nearly match that of the GRU if proper initialization is performed.

Bidirectional RNNs (BRNN)~\cite{schuster97} are another extension of the standard unidirectional RNNs
in which two RNNs are applied to the input data. Firstly, an
 RNN is applied on the input sequence which is then followed by the application of RNN
on the reversed input sequence.
This typically improves the accuracy of the model~\cite{baldi99,wei19}
at a cost of slower training times.~\cite{siaminamini19,wei19}
Several variants of bidirectional RNNs exist differing in the type of the underlying RNN cell.
Bidirectional LSTMs (BLSTM)~\cite{graves13} and bidirectional GRUs (BGRU)~\cite{lynn19} 
are being explored in various tasks.~\cite{graves13,tang16,ma17,kim19,wei19,zhang20}
In some applications,~\cite{siaminamini19,cui19} 
such as speech recognition, the better performance of BLSTM compared to
the regular unidirectional LSTM has been reported~\cite{graves13} 
and is not surprising given the nature of the task
(text parsing and prediction of next
words in the input sentence). In general, however, it is
not clear what problems benefit from the bi-directional training.


Convolutional neural networks (CNNs)~\cite{lecun89} 
are a special kind of ANNs designed for processing data that has a known grid-like topology.~\cite{goodfellow2016deep} For example, time-series data
can be thought of as a one-dimensional (1D) grid taking samples at regular time intervals. One-dimensional CNNs (1D CNNs) 
have achieved promising results in time-series
classification tasks~\cite{wang16,cui16,serra18,zheng14,zheng16,interdonato18,kashiparekh19,fawaz20,tang21} 
in many domains including healthcare,~\cite{rajpurkar17,roy18,schirrmeister18}
speech recognition,~\cite{sercu16,xiong17} music classification,~\cite{choi16,you18} 
natural language processing~\cite{yin17} and others.

CNNs have also been combined with RNNs. Such hybrid models are called
convolutional recurrent networks (CRNNs). A CRNN is a deep ANN that contains a CNN layer(s) followed by an 
RNN layer(s). Such architectures possess several advantages.
Firstly, 1D CNN layers learn to sub-sample the data and 
reduce the input vector that is passed to an RNN layer(s). 
This is important because GRU and LSTM layers
are computationally expensive and replacing some of them with convolutional layer(s)
improves the computational scaling of the algorithm. Second, 1D CNN layers extract local
information from neighboring time points and pass already detected
temporal dependencies further down to RNN layers.
CRNNs are being actively explored in various 
tasks.~\cite{karim18,roy18,you18,choi16,you18,sheykhivand20}
It has been reported that CRNNs can outperform CNNs in some tasks.~\cite{choi16}
CNNs were combined with BLSTMs as well.~\cite{eapen19} 
Note that these methods are different from convolutional LSTM 
models~\cite{shi15} in which
input transformations and recurrent transformations are both convolutional.

Kernel methods represent another class of ML methods that are applied to time-series analysis.~\cite{muller97,sapankevych09,haworth14}
Such methods employ a function (kernel) that maps input data
into a high dimensional space and then perform a linear regression in that space.
Kernel Ridge Regression (KRR) and Support Vector Regression (SVR) are examples of such algorithms.
A crucial aspect of applying kernel methods to
time series data is to find appropriate kernels to distinguish between time series. A
simple way is to treat time series as static vectors, essentially as they are treated in a feed-forward ANN,
ignoring the time dependence. In such cases standard kernels such as 
 Mat\'ern~\cite{gneiting10} or Gaussian radial basis function (RBF)~\cite{gneiting10} can be used. 
However, such methods are limited to input time-sequences of equal length, again similar to ANNs,
yet many applications
involve time sequences of varying length.
To overcome this limitation, 
kernel functions such as autoregressive~\cite{cuturi11} 
kernel have been developed. 


Recently, many ML models have been applied to simulate the dynamics of quantum systems.~\cite{rodriguez21,
ullah21,ullah2022one,ullah2022predicting,akimov21,secor21,yang20,bandyopadhyay18,banchi18,wu21,lin21,choi22,tsai22,tang22} 
We note that ML can also be applied to quantum dynamics in a different context---namely as surrogate models for quantum chemical properties such as 
potential energies and forces in different electronic states as well as couplings between states eliminating the need for expensive (excited-state) electronic structure calculations.~\cite{dral21,westermayrmlst21,westermayrcr22}
Here we apply ML to propagate a quantum system assuming potential energies are readily available.
ML models are attractive because of their very low computational cost and favorable scaling with respect
to the size of a quatum system and bath.
RNNs were used to simulate dynamics of the spin-boson model Hamiltonian and Landau--Zener transitions,~\cite{bandyopadhyay18,yang20}
and to learn the convolutionless master equation.~\cite{banchi18}
Rodriguez \textit{et al.}~\cite{rodriguez21} used CNNs to accurately model long-time dynamics of the spin-boson system
based on short-time dynamical data. Later Ullah \textit{et al.}~\cite{ullah21} illustrated that KRR methods can also predict
long-time dynamics of the spin-boson model very accurately. Recently, CNNs is used to study the excitation energy transfer in  Fenna--Matthews--Olson light-harvesting complex.~\cite{ullah2022predicting,ullah2022one}
Wu \textit{et al.}~\cite{wu21} used a hybrid CNN/LSTM network
to predict long-time semiclassical and mixed quantum-classical dynamics of the spin-boson model.
Lin \textit{et al.}~\cite{lin21,lin2022auto} trained a multi-layer LSTM model to simulate the long-time dynamics of 
spin-boson model and used bootstrap method to estimate the confidence interval.
We note that ML methods based on FFNNs have also been recently applied to model
quantum dynamics.~\cite{akimov21,secor21}
The recent upsurge of applications of ML methods to dissipative quantum dynamics 
calls for a systematic benchmark of such methods.

In this article, we present a comprehensive comparison of 22 ML models
for predicting the long-time dynamics of an open quantum system given the short-time evolution
data. We consider all three most used types of unidirectional RNNs including the vanilla
 RNN, GRU, and LSTM,
the corresponding bidirectional RNNs (BRNN, BGRU, BLSTM),  
1D CNNs, CRNNs, as well as KRR with 8 different kernel functions.
We test the ability of these ML methods to predict the donor-acceptor population difference
of a spin-boson model in symmetric and asymmetric regimes over a broad range of
temperatures, reorganization energies, and bath relaxation timescales. 

\section{\label{sec:theory}Theory}

\subsection{\label{sec:sb}Model system}
We choose to test ML methods on the spin-boson model
which has become a paradigmatic model system in the study of open quantum systems
due to the richness of its physics.~\cite{leggett87}
The spin-boson model was exploited in a wide-range of applications in quantum computing,~\cite{makhlin01} 
quantum phase transitions,~\cite{alvermann09,winter09}
electron transfer in biological systems,~\cite{garg85} and others. 
The spin-boson model
describes a two-level quantum subsystem linearly coupled to a heat-bath environment. The bath is modeled as an ensemble of 
independent harmonic oscillators. The total Hamiltonian in the subsystem's basis $\{|+\rangle, |-\rangle\}$ is given by ($\hbar=1$)
\begin{equation}
    \hat{H} = \frac{\epsilon}{2} \hat{\sigma}_z + \frac{\Delta}{2} \hat{\sigma}_x + 
    \hat{\sigma}_z \sum_\alpha g_\alpha \left(b_\alpha^\dagger + b_\alpha\right)
    + \sum_\alpha \omega_\alpha b_\alpha^\dagger b_\alpha,
\end{equation}
where $b^\dagger_\alpha$ ($b_\alpha$) is the bosonic creation (annihilation) operator
of the $\alpha$th mode with frequency $\omega_\alpha$,
$\hat{\sigma}_z=|+\rangle \langle+|-|-\rangle\langle-|$ and $\hat{\sigma}_x=|+\rangle \langle-|+|-\rangle\langle+|$ 
are the Pauli operators, $\epsilon$ is the energetic bias, 
$\Delta$ is the tunneling matrix element, and $g_\alpha$ are the coupling coefficients. 

The  impact  of  the  bath  is  completely  determined  by the spectral density
$J(\omega)=\pi \sum_\alpha g_\alpha^2 \delta(\omega_\alpha - \omega)$
which, in this work, is choosen to be of the Debye form (Ohmic spectral density with the Drude--Lorentz cut-off)~\cite{wang99}
\begin{equation}
    J(\omega)=2\lambda \frac{\omega \omega_c}{\omega^2 + \omega_c^2},
\end{equation}
where $\lambda$ is the bath reorganization energy which controls the strength of the coupling between system and the bath, 
and the cutoff frequency $\omega_c$ which sets the primary timescale for the bath evolution $\tau_c = (\omega_c)^{-1}$.

We consider the time evolution of the expectation value of $\hat{\sigma}_z$ Pauli operator 
\begin{equation}
    \langle {\sigma}_z(t)\rangle= \mathrm{Tr}_\mathrm{s} \left[ \hat{\sigma}_z \hat{\rho}_\mathrm{s}(t)\right], \label{eq:sbz}
\end{equation}
which is often referred to as the population difference $\langle {\sigma}_z(t)\rangle=p_+(t)-p_-(0)$, where
$p_\pm(t) = \mathrm{Tr}_\mathrm{s} \left[  |\pm\rangle \langle \pm| \hat{\rho}_s(t)\right]$.
In Eq.~\eqref{eq:sbz} the trace is taken over the system degrees of freedom as denoted by ``s'', and $\hat{\rho}_\mathrm{s}$
is the system's reduced density operator
\begin{equation}
 \hat{\rho}_\mathrm{s}(t) = \mathrm{Tr}_\mathrm{b} \left[ e^{-i\hat{H}t} \hat{\rho}(0) e^{i\hat{H}t} \right],
\end{equation}
where $\hat{\rho}(0)$ is the total system plus bath density operator and the trace is taken over the bath 
degrees of freedom. The initial state of the total system is assumed to be a product state of the following form
\begin{equation}
    \hat{\rho}(0) = \hat{\rho}_\mathrm{s}(0) \otimes \frac{e^{-\beta \hat{H}_\mathrm{b}}}{Z_\mathrm{b}},
\end{equation}
where $Z_\mathrm{b}=\mathrm{Tr}_\mathrm{b}\left[e^{-\beta \hat{H}_\mathrm{b}}\right]$ 
is the bath partition function, $\beta=(k_\mathrm{B}T)^{-1}$ is the inverse temperature, and $k_\mathrm{B}$ is the
Boltzmann constant. The initial density operator of the system is chosen to be $\hat{\rho}_\mathrm{s}(0)=|+\rangle \langle +|$.
These conditions correspond to situations where the initial preparation of the subsystem occurs 
quickly on the timescale of the bath relaxation.

\subsection{\label{sec:methods}Machine learning models}
In this section we provide a detailed description of all ML models used in
the present study. We specialize our discussion to modeling time-series data with the input 
sequence denoted as $\mathbf{x}=\left(x^{(1)},\ldots,x^{(T)}\right)$ where $T$ is the length of
the time series.
Each element $x^{(t)}$ of $\mathbf{x}$ can be a real-valued vector itself, $x^{(t)} \in \mathbb{R}^n$.
In this work, the dimension $n$ of each element of an input sequence $\mathbf{x}^{(t)}$ is 1.
Consider a data set $\mathcal{D}=\{(\mathbf{x}_i,\mathbf{y}_i)\}_{i=1}^N$
containing $N$ time series $\mathbf{x}_i$ and their associated labels  $\mathbf{y}_i$. In time-series forecasting problems,
labels can describe the future states of the input sequence $\mathbf{x}_i$ as denoted by
$\mathbf{y}_i=\left(x_i^{(T+1)},\ldots,x_i^{(T+m)}\right)$ which corresponds to $m$ next elements of the
sequence $\mathbf{x}_i$.
Time-series forecasting in a supervised learning framework amounts to (\textit{i}) training 
ML models on the subset of $\mathcal{D}$ called a training set
and (\textit{ii}) using trained ML models to make a prediction  
$\hat{\mathbf{y}}_i$ for a given test time series $\mathbf{x}_i$.
In this work we test the ability of ML models to predict a single real-valued
scalar quantity, the population difference of the
spin-boson model,
$\langle {\hat{\sigma}}_z(t)\rangle$ for a single time step i.e., $m=1$. 
Extension to multivariate time series data ($n > 1$) and multi-step outputs ($m > 1$) is possible.\cite{ullah2022one}

\subsubsection{\label{sec:ffnn} Feedforward neural networks}
Feedforward neural networks (FFNNs) or multiplayer perceptrons 
are the most used type of artificial neural networks. FFNN approximate a function $g(\mathbf{x})$ by
defining a mapping $\mathbf{y}=\mathcal{F}(\mathbf{x};\boldsymbol{\theta})$
and determining the value of parameters $\boldsymbol{\theta}$ that best approximate $g(\mathbf{x})$. These models are called
feedforward because information flows through the function being evaluated from an input $\mathbf{x}$ through
intermediate steps to the output $\hat{\mathbf{y}}$. There are no feedback connections between the output and the input of the model.~\cite{goodfellow2016deep}

A crucial element of the success of ANNs is the use of deep architectures. Deep ANNs are created
by stacking multiple layers on top of each other, with the output of one layer forming the
input sequence for the next layer. The first layer is
called an input layer, the last layer is called the output layer; all layers in between are called hidden layers.

Deep FFNNs are compositions of several functions: $\hat{\mathbf{y}}=\mathcal{F}^{(L)}\large(\ldots \mathcal{F}^{(2)}(\mathcal{F}^{(1)}\left(\mathbf{x};\boldsymbol{\theta}^{(1)})\right);$ $\boldsymbol{\theta}^{(2)});\boldsymbol{\theta}^{(L)}\large)$ 
where each function $\mathcal{F}^{(k)}$
depends on its own set of parameters $\boldsymbol{\theta}^{(k)}$. Function $\mathcal{F}^{(l)}$
connects $l$th and $(l+1)$th layers of the network. It takes an input $\mathbf{x}^{(l)}\in \mathbb{R}^{k_l}$ 
and generates the output  according to
\begin{equation}
    \mathbf{x}^{(l+1)} = \mathcal{F}^{(l)}\left(\mathbf{x}^{(l)};\boldsymbol{\theta}^{(l)}\right)=f^{(l)} \left( \mathbf{a}^{(l)}\right), \label{eq:ffnn}
\end{equation}
where $f^{(l)}:\mathbb{R}\to \mathbb{R}$ 
is the $l$th layer activation function which is applied elementwise. As seen from Eq.~\eqref{eq:ffnn} each 
layer of the network is vector valued. Each vector element $a_j^{(l)}$ is called a neuron. Its value is calculated from layer's input and parameters 
\begin{equation}
a_j^{(l)}=\sum_{i=1}^{k_l} w_{ij}^{(l)}x_i^{(l)} + b_j^{(l)},    
\end{equation}
where $\mathbf{w}^{(l)}\in\mathbb{R}^{k_l \times k_{l+1}}$ and $\mathbf{b}^{(l)}\in\mathbb{R}^{k_l}$ are called the weights and the biases of the $l$th 
layer, respectively, 
that together constitute the set of
trainable parameters of the $l$th layer $\boldsymbol{\theta}^{(l)}=\{\mathbf{w}^{(l)},\mathbf{b}^{(l)}\}$. The output $\mathbf{x}^{(l+1)}$ 
forms an input into $(l+1)$th layer.  The total number of trainable parameters, biases and weights, 
of a single FFNN layer is $k_l(k_{l+1} + 1)$. It should be emphasized that FFNN models require input time sequences of a fixed 
length, $T$. This, in particular,
requires 
an \textit{a priori} 
knowledge of the memory effects in a quantum system under study which is non-trivial task. Artificial neural networks based on recurrent layers, 
discussed below,
do not impose such a restriction.


The strategy of deep learning is to find the set of model parameters $\{\boldsymbol{\theta}^{(1)},\ldots,\boldsymbol{\theta}^{(L)}\}$
that best approximates the target function $g(\mathbf{x})$. The model parameters are adjusted during the training process which 
is based on backpropagation algorithm.~\cite{goodfellow2016deep}

Typically the activation function of the hidden layers is chosen to be nonlinear, e.g., logistic sigmoid function
$\sigma(z)=(1+e^{-z})^{-1}$. 
According to the universal approximation theorem~\cite{hornik89,cybenko89,leshno93} an FFNN with a linear output activation function and at least one hidden layer with a nonlinear activation function can approximate any Borel measurable function (continuous on the closed and bonded subset of the real
coordinate space) 
from one finite-dimensional
space to another with any desired nonzero amount of error, provided the network contains enough neurons. The theorem gurantees that
regardless of the target function, a single hidden layer FFNNs with many neurons will be able to represent this function with any degree of accuracy. 
However, in general, there is no
guarantee that the training algorithm can do so.~\cite{goodfellow2016deep}

\subsubsection{\label{sec:1dcnn}1D convolutional neural networks}
Convolutional neural networks (CNNs) is a type of ANNs 
that can be applied to time-series data.~\cite{lecun89,goodfellow2016deep}  
CNNs are based on a mathematical operation called convolution.
Let $\left(\mathbf{g} \ast \mathbf{w}\right)$ be the result of a 1D discrete convolution and 
the $i$th element of the result is given by
\begin{equation}
\left(\mathbf{g} \ast \mathbf{w}\right)_i = \sum_{j=1}^k g_{i+j-1} w_{k+1-j},    \label{eq:cnnt}
\end{equation}
where $\mathbf{w} \in \mathbb{R}^{k}$ is referred to as kernel, or weight vector, and $\mathbf{g}$ is a time series.
CNNs exploit two key ideas: sparse connectivity and parameter sharing.~\cite{goodfellow2016deep} 
The former is accomplished by making the kernel size smaller than the size of the input  which allows to detect, for the time-series data, 
short-time correlations and reduces the number of parameters to be stored in  memory
 compared to FFNNs. The latter amounts to using the same kernel parameters for all 
positions in the input which further reduces the storage requirements. In the case of convolution, the parameter
sharing leads to equivariance---a property that causes the output to change in the same way the input changes.
Specifically, in the case of time-series data, the convolution captures a timeline for different features to
appear in the input.~\cite{goodfellow2016deep} 

In practical implementations of CNNs, 
the convolution operation given in Eq.~\eqref{eq:cnnt} is often replaced by the
so-called cross-correlation operation~\cite{goodfellow2016deep} which we denote by a ``$\star$''
\begin{equation}
    \left(\mathbf{g} \star \mathbf{w}\right)_i = \sum_{j=1}^{k}   g_{i+j-1} w_{j}.\label{eq:cc}
\end{equation}
It should be noted that Eq.~\eqref{eq:cc} is given for 
the step size of the kernel, as it is applied across the sequence, equal to 1. 
This step size is called stride $s (s\in \mathbb{Z}^+)$. In general, for the stride $s \ge 1$, Eq.~\eqref{eq:cc} 
is modified to
\begin{equation}
    \left(\mathbf{g} \star \mathbf{w}\right)_i = \sum_{j=1}^{k}   g_{(i-1)s+j} w_{j}.\label{eq:ccs}
\end{equation}

Let's consider a two-layer 1D CNN architecture. 
An input sequence $\mathbf{x}$ 
is processed through $K_1$ 1D kernels of the first layer whose weights are denoted as
$\mathbf{w}^{(1)} \in \mathbb{R}^{k_1}$. The result of this operation is a tensor $\mathbf{Z}^{(1)}$
whose elements are given by
\begin{equation}
    Z_{ij}^{(1)} = f^{(1)} \left[ \left( \mathbf{x} \star \mathbf{w}^{(1)}_j \right)_i + b_j^{(1)} \right], \label{eq:ccz}
\end{equation}
where $f^{(1)}$ is the activation function,
$\mathbf{b}^{(1)} \in \mathbb{R}^{K_1}$ are the bias parameters and $\mathbf{w}_j^{(1)},j=1,\ldots,K_1$ is the $j$th kernel
of the first layer.
The output  tensor 
$\mathbf{Z}^{(1)}$ has dimensions of $M_1\times K_1$ with $M_1 = \lfloor (T - k_1 + 2p_1)/s_1 \rfloor  + 1$, where
$s_1$ is the stride, 
$p_1$ is the amount of zero padding, and $\lfloor x \rfloor = \max \{ m \in \mathbb{Z} | m \le x\}$ 
is the floor function. 
Zero padding parameter $p_1$ denotes the number of zeros to be added to the start 
and the end of the input sequence. 
For example, for $p=0$ the input sequence is not padded with zeros and, consequently, the kernel is allowed to visit only
positions where it is contained entirely within the input sequence. This type of zero padding is called a valid padding.

Because convolution of input data with a single kernel can extract only one kind of features, albeit at many locations,
in practice, many filters are used.
Each element of a kernel is a trainable parameter and the
same parameters of each kernel are shared across the entire sequence. 
Therefore, the total number of trainable parameters of the first 1D CNN layer is $K_1(k_1+1)$.

The output of the first layer $\mathbf{Z}^{(1)}$ is then processed through $K_1 \times K_2$ 1D
kernels whose weights are denoted as $\mathbf{w}^{(2)}\in \mathbb{R}^{k_2}$. The output of the second layer
is tensor $\mathbf{Z}^{(2)}$ whose elements are given by
\begin{equation}
    Z_{ij}^{(2)} = f^{(2)} \left[ \sum_{u=1}^{K_1} 
    \left( \mathbf{Z}^{(1)}_u \star \mathbf{w}_{uj}^{(2)} \right)_i + b_j^{(2)}\right],\label{eq:cnn2}
\end{equation}
where $f^{(2)}$ is the activation function of the second layer and $\mathbf{b}^{(2)}$ are the bias
parameters of the second layer. 
In Eq.~\eqref{eq:cnn2} $\mathbf{w}_{uj}^{(2)}$ is to be understood as the
 the $u$th kernel (weight vector) applied to the output of $j$th kernel (also called channel) 
of the first layer. 
The dimensions of 
$\mathbf{Z}^{(2)}$ are $M_2\times K_2$ where $M_2 = \lfloor (M_1 - k_2 + 2p_2)/s_2 \rfloor  + 1$
with $p_2$ and $s_2$ being the zero-padding and the stride of the second 1D CNN layer, respectively.
It is worth emphasizing that a separate set of kernels is applied to the output
of each kernel of the preceding (in this case first) layer. Thus,
the total number of trainable parameters of the second 1D CNN layer is $K_2(K_1k_2+1)$.

A pooling layer is commonly found in CNN architectures. 
Its function is to subsample (shrink) the input.  
Pooling function replaces the output of a CNN layer at a certain location with a neighborhood-dependent
information. For example, the maximum  pooling (MaxPooling) operation~\cite{zhou88} 
outputs the maximum value in a  neighborhood of a specified size.
This helps to make representation approximately invariant to small translations of the input and improves the computational
efficiency in a deep CNN by reducing the number of trainable parameters of the next convolutional layer.
A 1D MaxPooling layer with the kernel size $k_{p}$, stride $s_{p}$, 
operates independently on every depth slice of the input $\mathbf{Z}^{\mathrm{in}}$ and resizes it 
by taking only the maximum value 
\begin{equation}
    Z_{ij}^{\mathrm{out}} = \max \left( Z_{(i-1)s_{p}+1,j}^{\mathrm{in}},\ldots, Z_{(i-1)s_{p}+k_{p},j}^{\mathrm{in}} \right),\label{eq:mp}
\end{equation}
where $i=1,\ldots,\lfloor (M_2 - k_p)/s_p\rfloor+1$. 
Note that Eq.~\eqref{eq:mp} is written, for simplicity, for the zero padding
but other kinds of padding can be used as well.
MaxPool layers do not employ any trainable parameters.


\subsubsection{\label{sec:rnn}Recurrent neural networks}
Recurrent neural networks (RNNs) are a family of neural networks specifically designed for processing sequential data.~\cite{rumelhart86} Similarly to
FFNNs, RNNs have the universal approximation ability.~\cite{schafer07,hornik89}
Similarly to CNNs, RNNs are based on the idea of parameter sharing but, unlike CNNs, RNNs share parameters through 
recursion. Formally, given a sequence $\mathbf{x}$ at each time step, an RNN updates its hidden state $\mathbf{H}=\left[ \mathbf{h}^{(1)},\ldots,\mathbf{h}^{(T)}\right]$
recursively
 based on the current input $x^{(j)}$ and the previous hidden state
$\mathbf{h}^{(j-1)}$ as follows
\begin{equation}
\mathbf{h}^{(j)}=\mathcal{R}(\mathbf{h}^{(j-1)},x^{(j)}), \label{eq:frnn} 
\end{equation}
where $\mathcal{R}$ is a nonlinear function.
Each element of the hidden state vector, $\mathbf{h}^{(j)}$ is, in general, a vector itself.
Recursive application of Eq.~\eqref{eq:frnn} results in the sharing of parameters across an ANN architecture.
Training of such ANN amounts to 
selectively emphasizing some aspects of the past sequence inputs 
that deemed more important than others. 

\begin{figure}
    \centering
    \includegraphics[width=0.5\textwidth]{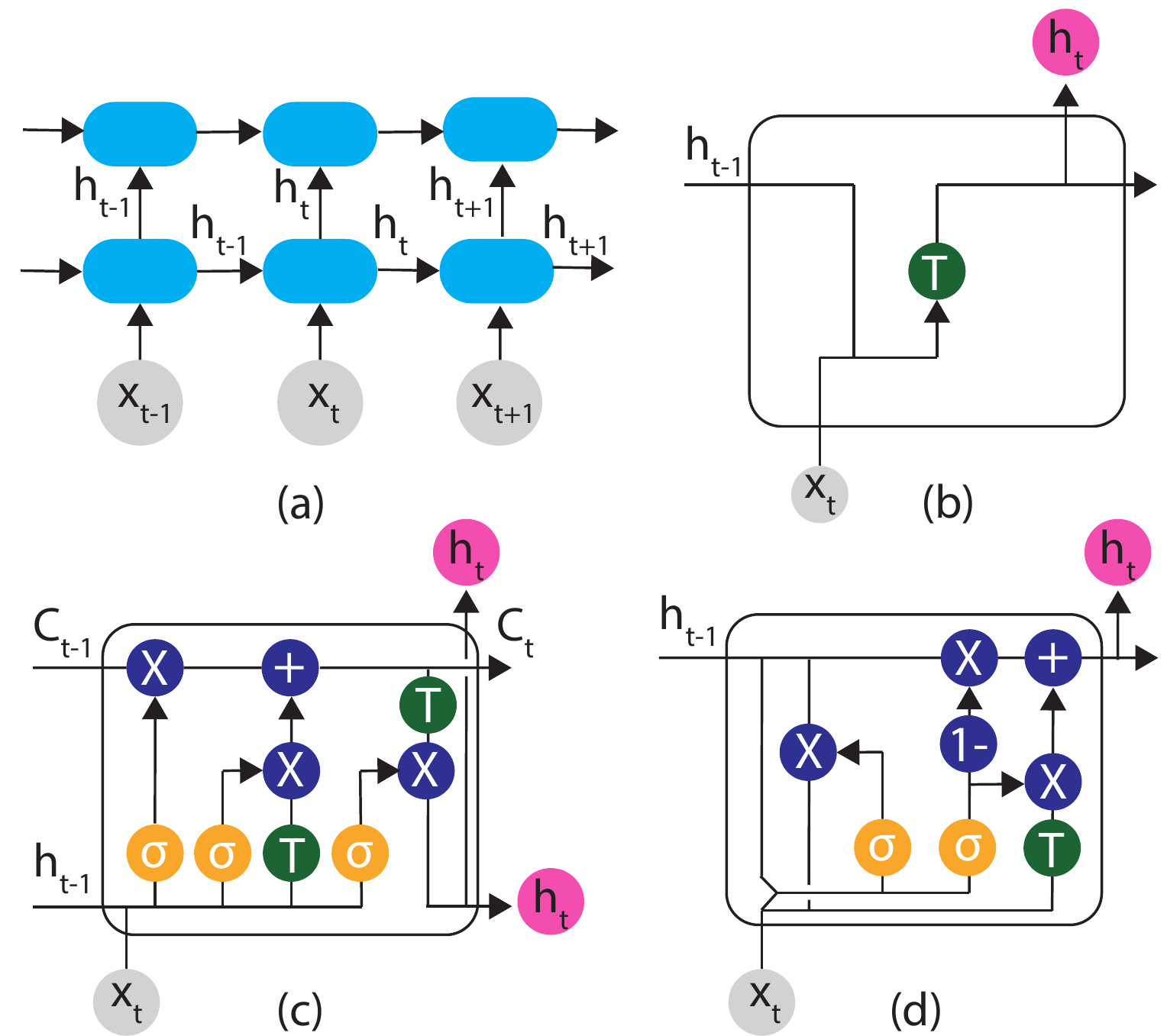}
    \caption{(a) Basic structure of a two-layer unidirectional
    recurrent neural network. Blue rounded rectangles represent cells and $x_t$
    is the $t$th element of an input sequence $\mathbf{x}$. 
    Hidden state vector $\mathbf{h}^{(t)}$ is passed between the 
    cells within a layer as well as
    between the layers; (b) the vanilla RNN cell with \textit{tanh} activation function; (c) LSTM cell; (d) GRU cell. Green and yellow circles denote
    hyperbolic tangent and sigmoid activation functions correspondingly. Dark blue circles denote
    element-wise arithmetic operations shown on the circle. See text for details.}
    \label{fig:fig1}
\end{figure}

All RNN models used in this work contain two stacked recurrent 
layers and are schematically illustrated in Fig.~\ref{fig:fig1}a. 
The number of RNN cells in the first layer equals to the length of an input vector such that
each cell processes a single time step $t=1,\ldots,T$ and updates the corresponding
hidden state vector $\mathbf{h}^{(t)} \in \mathbb{R}^{k_1}$.
The size of the hidden state vector $k_1$ 
is a hyperparameter that needs to be optimized. 
Note that various software packages use different terminology.
In this work RNN models are built with the \textsc{Keras}~\cite{keras} software whose argument \textsc{units} corresponds to the
size of the hidden state vector. An RNN layer outputs a tensor, $\mathbf{H} \in \mathbb{R}^{T \times k_1}$, 
containing all hidden
states 
and it serves as an input to the following RNN layer. 

Each RNN cell of a layer processes the corresponding slice of the preceding layer's output
$\mathbf{H}_{Lj}$, where $j \in [1,k_1]$ and $L$ is the number of cells of the second layer,
and transforms it into the state vector $\mathbf{h}_2^{(t)} \in \mathbb{R}^{k_2}$,
where $k_2$ is the number of units of the second RNN layer. 
The last RNN layer can either return 
  the hidden state vector at only the final time step $T$ or the entire sequence of hidden states. 

Various RNN architectures differ 
by the function $\mathcal{R}$ in Eq.~\eqref{eq:frnn}.
A simple RNN cell, also known as vanilla RNN is
illustrated in Fig.~\ref{fig:fig1}b with the hyperbolic tangent activation function.
It receives an input $x^{(t)}$ corresponding
to the time step $t$ and uses the state vector from the chronologically 
previous RNN cell $\mathbf{h}^{(t-1)} \in \mathbb{R}^k$ to generate the hidden state $\mathbf{h}^{(t)}$
according to the following update equation
\begin{equation}
    \mathbf{h}^{(t)} = \tanh\left(\mathbf{b} + \mathbf{w}_{h} \cdot \mathbf{h}^{(t-1)} + \mathbf{w}_{x} \cdot \mathbf{x}^{(t)}\right),\label{eq:rnn}
\end{equation}
where $\mathbf{b}_h \in \mathbb{R}^{k}$ are the bias parameters, $\mathbf{w}_h \in \mathbb{R}^{k \times k}$,
$\mathbf{w}_x \in \mathbb{R}^{k \times m}$ are the
coefficient matrices, and  the \textit{tanh} activation function
 is applied element-wise.
The function $\mathcal{R}(\mathbf{h}^{(j-1)},x^{(j)})$ in case of the vanilla RNN
amounts to iterative application of Eq.~\eqref{eq:rnn} to an input sequence
$\mathbf{x}$ from $t=1$ to $t=T$.
 
The hidden state $\mathbf{h}^{(t)}$ is then passed to the next RNN cell. 
Weights $\mathbf{w}_h,\mathbf{w}_x$ and biases $\mathbf{b}$ in Eq.~\eqref{eq:rnn} 
are updated iteratively via the back-propagation
algorithm. In a multilayer RNN architecture all hidden state vectors 
$\mathbf{H}$ are passed to the next layer.
For RNNs whose output is used directly to predict the next value of a sequence
the hidden state corresponding to $t=T$ 
is the desired predicted value of the next time step $\hat{\mathbf{y}}=\mathbf{h}^{(T)}$.

The total number of trainable parameters of the vanilla RNN cell
is $k(k+n+1)$, where $k$ is the size of the hidden state vector, which is the same for all RNN cells within a layer, and
$n$ is the dimension of each element of an input sequence $\mathbf{x}^{(t)}$
which, in this work is 1 for the first RNN layer and $k_1$ for the second RNN layer. Note that if RNN is not the first layer,
as e.g., in convolutional recurrent neural networks discussed below, $n$ would be different from 1.
It should also be noted that the total number of trainable parameters of the whole RNN layer 
comprised of any number of the vanilla RNN cells is the same because $\mathbf{b}, \mathbf{w}_h$, and $\mathbf{w}_x$ are 
shared across all RNN cells within the given layer. The total number of trainable parameters in a two-layer vanilla
RNN architecture
is $k_2(k_1+k_2+1)+k_1(k_1+1+1)$. 

In spite of being simple and powerful ANN model, the vanilla RNN described above
is plagued by the vanishing gradient and exploding gradient problems when trained 
via backpropagation.~\cite{goodfellow2016deep,hochreiter91,bengio94,bengio93} 
The exploding gradients problem refers to the large increase 
in the norm of the gradient during training. The vanishing gradients problem refers to the opposite behavior, 
when long term components decay exponentially fast to zero
norm, limiting the model's ability to learn long-range dependencies.
The exploding gradient problem
can be easily addressed by gradient clipping.~\cite{pascanu13}
The vanishing gradient problem is, however, much more difficult to address. 

\subsubsection{\label{sec:lstm}Long short-term memory networks}
In general, learning long-term dependencies is one of the most important problems in deep learning.~\cite{goodfellow2016deep}
The difficulty in dealing with long-memory sequences in RNNs arises from the exponentially
smaller weights given to long-term correlations compared to short-term ones.~\cite{goodfellow2016deep,hochreiter91,bengio94,bengio93}
This problem is particular to simple RNN cell described above. It was shown that very deep FFNNs 
can avoid the vanishing and exploding gradient problems~\cite{sussillo14} but, in order to
store memories,  RNNs must enter a region of parameter space where gradients vanish.~\cite{bengio93,bengio94}

To overcome the vanishing gradient problem various gated RNN architectures such as long 
short-term memory (LSTM) were developed. 
Gated RNNs are based on the idea of creating paths
through time such that the derivatives are neither vanish nor explode. 
The Long Short-Term Memory (LSTM)~\cite{hochreiter97,graves05,gers00,gers99} model 
was developed by Hochreiter \textit{et al.}~\cite{hochreiter97} and is based 
on the idea of using self-loops
to produce paths where the gradient can flow for long duration. This 
is achieved by adding the cell state denoted by $\mathbf{C}^{(t)}\in \mathbb{R}^k$
and data-dependent gates, that control the flow of information, to the standard RNN architecture.~\cite{hochreiter97,gers00} 
Both hidden state and cell state control the memory of the network.
The cell state carries relevant information throughout the processing of the sequence
such that even information from the earlier time steps can make its way to later time steps, 
reducing the effects of short-term memory. The information is added or removed to the cell state via gates. 
The gates are different simplest FFNNs that decide which information is allowed on the cell state. 
Thus, the gates can learn what information is relevant to keep or forget during training.

Although a number of variants of the LSTM cell have been produced, a large-scale analysis shows that none of them
outperforms the standard LSTM architecture.~\cite{hochreiter97,graves05,gers00}
The block diagram of the LSTM cell is shown in Fig.~\ref{fig:fig1}c. 
It contains three gates: the forget gate, the input gate, and the output gate. 
All gates have a sigmoid [logistic $\sigma(z)=(1+e^{-z})^{-1}$],
nonlinear activation function
(yellow circles). The t$th$ element(s) of the input sequence $\mathbf{x}^{(t)}$ enters the cell and is concatenated
with  the hidden state  vector  from  the  chronologically previous cell $\mathbf{h}^{(t-1)}$. The total vector is
passed through the forget gate 
\begin{equation}
    \mathbf{F}^{(t)} = \sigma \left( \mathbf{b}^F + \mathbf{W}_h^F \cdot \mathbf{h}^{(t-1)} + 
    \mathbf{W}_x^F \cdot \mathbf{x}^{(t)}\right),
\end{equation}
where $\mathbf{b}^F \in \mathbb{R}^k$ are the biases and $\mathbf{W}_h^F \in \mathbb{R}^{k \times k}$, 
$\mathbf{W}_x^F \in \mathbb{R}^{k \times m}$,
are the corresponding weights. The forget gate was the crucial addition to the original LSTM cell
extending the length of sequences that can be processed.~\cite{gers00} The forget gate
learns to reset memory blocks once their contents are out of date and hence
useless.
The output vector $\mathbf{F}^{(t)} \in \mathbb{R}^k$ contains the values between 1 and 0  
emphasizing or diminishing the importance of the elements of the hidden state vector, correspondingly. 
In two separate branches, the current input vector and the previous hidden state
vector are processed through the input gates: the external input gate 
\begin{equation}
    \mathbf{I}^{(t)} = \sigma \left( \mathbf{b}^I + \mathbf{W}_h^I \cdot \mathbf{h}^{(t-1)} + \mathbf{W}_x^I \cdot \mathbf{x}^{(t)}\right),
\end{equation}
and the new candidate gate
\begin{equation}
    \mathbf{G}^{(t)} = \tanh \left( \mathbf{b}^G + \mathbf{W}_h^G \cdot \mathbf{h}^{(t-1)} + \mathbf{W}_x^G \cdot \mathbf{x}^{(t)}\right),
\end{equation}
where $\mathbf{b}^I \in \mathbb{R}^k$, $\mathbf{b}^G \in \mathbb{R}^k$, $\mathbf{W}_h^I \in \mathbb{R}^{k \times k}$, 
$\mathbf{W}_h^G \in \mathbb{R}^{k \times k}$, 
$\mathbf{W}_x^I \in \mathbb{R}^{k \times m}$, and $\mathbf{W}_x^G \in \mathbb{R}^{k \times m}$ are the corresponding biases 
and weights of the input and new candidate gates. The external
output gate uses the sigmoid activation function to obtain a gating value between 0 and 1 which is then
element-wise multiplied by the corresponding element of the new candidate vector $\mathbf{G}^{(t)}$ effectively
deciding which elements of the input vector are the most important. Given $\mathbf{F}^{(t)}$, $\mathbf{I}^{(t)}$, and
$\mathbf{G}^{(t)}$ the cell state of the previous time step $\mathbf{C}^{(t-1)}$ is updated as follows
\begin{equation}
    \mathbf{C}^{(t)} = \mathbf{F}^{(t)} \odot \mathbf{C}^{(t-1)} + \mathbf{I}^{(t)} \odot \mathbf{G}^{(t)}, \label{eq:lstmc}
\end{equation}
where $\odot$ denotes the element-wise vector (Hadamard) product. The first term in Eq.~\eqref{eq:lstmc} 
removes parts of the previous cell state through the forget gate and the second term 
adds new information yielding the new cell state.

Finally, the hidden state $\mathbf{h}^{(t)}$ 
is updated as follows
\begin{eqnarray}
    \mathbf{O}^{(t)} & = &
    \sigma \left( \mathbf{b}^O + \mathbf{W}_h^{O} \cdot \mathbf{h}^{(h-1)} + \mathbf{W}_x^{O} \cdot \mathbf{x}^{(t)} \right),\\
    \mathbf{h}^{(t)} & = & \mathbf{O}^{(t)} \odot \tanh \left( \mathbf{C}^{(t)} \right),
\end{eqnarray}
where $\mathbf{O}^{(t)}$ is known as the output gate which uses a sigmoid activation function
for gating and the corresponding bias
parameters $\mathbf{b}^O \in \mathbb{R}^k$ and weights $\mathbf{W}_h^{O} \in \mathbb{R}^{k \times k}$ and 
$\mathbf{W}_x^{O} \in \mathbb{R}^{k \times m}$. The new hidden state $\mathbf{h}^{(t)}$ 
is then used to compute what to forget, input, 
and output by the cell in the next time step. Thus, while both hidden state and
cell state control the memory of the network, the cell state carries information about the entire
sequence and the hidden state encodes the information about the most recent time step.
Similarly to a deep vanilla RNN architecture, in a deep LSTM NN the
hidden state of each LSTM cell is passed to the next layer, as shown in Fig.~\ref{fig:fig1}a.

Since each of the four gates are based on a perceptron 
the total number of trainable parameters
of the LSTM variant shown in Fig.~\ref{fig:fig1}c is $4k(k+m+ 1)$ and, since $\mathbf{b}^F, \mathbf{b}^I, \mathbf{b}^O$, 
$\mathbf{W}_h^F$, $\mathbf{W}_h^I$, $\mathbf{W}_h^O$, 
$\mathbf{W}_x^F$, $\mathbf{W}_x^I$, and $\mathbf{W}_x^O$ are  shared  across  all  LSTM  cells,
this is also the total number of trainable parameters of the first LSTM layer comprised of any number of 
LSTM cells. The total number of trainable parameters of a two-layer LSTM architecture is
$4 [k_2 (k_1 + k_2+1) +  k_1 (k_1 + m+1)]$.
Thus, such LSTM models 
contain exactly four times more trainable parameters than the vanilla RNN models with the same
number of \textsc{units} (length of the hidden state vector).

\subsubsection{\label{sec:gru}Gated recurrent unit networks}
LSTM  has become a popular off-the-shelf architecture that effectively solves the vanishing gradient problem.
However, LSTM is often criticized for its \textit{ad hoc} nature. Furthermore, the purpose of its many
components is not apparent and there is no proof that LSTM is even the optimal structure.
For example, Ref.~\onlinecite{jozefowicz15} shows that the forget gate
is crucial element of the LSTM architecture, while the output gate is the least important.

Gated recurrent unit (GRU)~\cite{cho14,chung14,chung15,jozefowicz15} ANNs are another example of gated RNNs. GRUs
were introduced as a simplified version of LSTM that uses one less gate and, thus, has fewer
trainable parameters. The main difference between LSTM and GRU cells is that 
forget and input gates appear in the LSTM architecture are combined together in one gate in the GRU cell.
Additionally, the hidden state and cell state are combined as well. 
The information between GRU cells is transferred via the hidden state vector.
Due to the reduction in trainable parameters
and complexity
models based on GRU cells tend to converge faster.

There exist several
variants of a GRU cell. Fig.~\ref{fig:fig1}d illustrates the GRU cell used in this work.
 The update equations are discussed below. An input vector $\mathbf{x}^{(t)}$ is concatenated
 with the previous cell hidden state $\mathbf{h}^{(t-1)}$ and passed through the update $Z$ and reset $R$ 
 gates both using the sigmoid activation function
 \begin{eqnarray}
     \mathbf{Z}^{(t)} & = & \sigma \left( \mathbf{b}^Z + \mathbf{W}_h^Z \cdot \mathbf{h}^{(t-1)} + 
    \mathbf{W}_x^Z \cdot \mathbf{x}^{(t)}\right), \\
    \mathbf{R}^{(t)} & = & \sigma \left( \mathbf{b}^R + \mathbf{W}_h^R \cdot \mathbf{h}^{(t-1)} + 
    \mathbf{W}_x^R \cdot \mathbf{x}^{(t)}\right),
 \end{eqnarray}
 where $\mathbf{b}^{Z(R)} = \mathbf{b}_{W_h}^{Z(R)} + \mathbf{b}_{W_x}^{Z(R)} 
 \in \mathbb{R}^k$ are the biases and $\mathbf{W}_h^{Z(R)} \in \mathbb{R}^{k \times k}$, 
$\mathbf{W}_x^{Z(R)} \in \mathbb{R}^{k \times m}$ are the weights of the update (reset) gate.
The update gate controls the update of the memory state $\mathbf{h}^{(t)}$. The reset gate
controls the influence of the hidden state $\mathbf{h}^{(t-1)}$ on $\mathbf{G}^{(t)}$ introducing
additional nonlinear effect in the relationship between past and future states. The two gates
can individually ``ignore'' parts of the hidden state vector.~\cite{goodfellow2016deep}

Next, the input vector $\mathbf{x}^{(t)}$
is concatenated with the element-wise product of the hidden state vector $\mathbf{h}^{(t-1)}$ and the
output of the reset gate which, after passing through hyperbolic tangent activation function, produces 
 the so-called proposed new candidate state
\begin{equation}
     \mathbf{G}^{(t)}  =  \tanh \left( \mathbf{b}^G + 
     \mathbf{W}_h^G \cdot \left( \mathbf{R}^{(t)} \odot \mathbf{h}^{(t-1)} \right) + 
    \mathbf{W}_x^G \cdot \mathbf{x}^{(t)}\right) ,
 \end{equation}
 where $\mathbf{b}^{G} = \mathbf{b}_{W_h}^{G} + \mathbf{b}_{W_x}^{G} \in \mathbb{R}^k$, 
 $\mathbf{W}_h^G \in \mathbb{R}^{k \times k}$, 
 and $\mathbf{W}_x^G \in \mathbb{R}^{k \times m}$ are the corresponding bias vector and weights. The hidden
 state is updated by combining the previous cell hidden state $\mathbf{h}^{(t-1)}$ with the output of the
 update gate as well as with the product of the output of the update gate and the proposed candidate gate
 \begin{equation}
     \mathbf{h}^{(t)} = \left( 1- \mathbf{Z}^{(t)} \right) \odot \mathbf{h}^{(t-1)} + \mathbf{Z}^{(t)} \odot \mathbf{G}^{(t)}.\label{eq:gruh}
 \end{equation}
The total number of trainable parameters of a single GRU cell described above,
as implemented in \textsc{TensorFlow} software library, 
is $3k(k+m+ 2)$. Note the use of the two separate sets of biases $\mathbf{b}_{W_h}$ and $\mathbf{b}_{W_x}$.
The total number of trainable parameters is independent of the number of GRU cells due to parameter sharing.
Two-layer GRU models contain $3 [k_2 (k_2 + k_1+2) +  k_1 (k_1 + m+2)]$ trainable parameters. Thus, given the same
$k_1$, $k_2$, and $m$, GRU
models contain approximately three times more parameters than the corresponding vanilla RNN models
and \textit{ca.} 1/4 less parameters than LSTM models with the same number of layers.

\subsubsection{\label{sec:brnn}Bidirectional RNNs}

\begin{figure}
    \centering
    \includegraphics[width=0.5\textwidth]{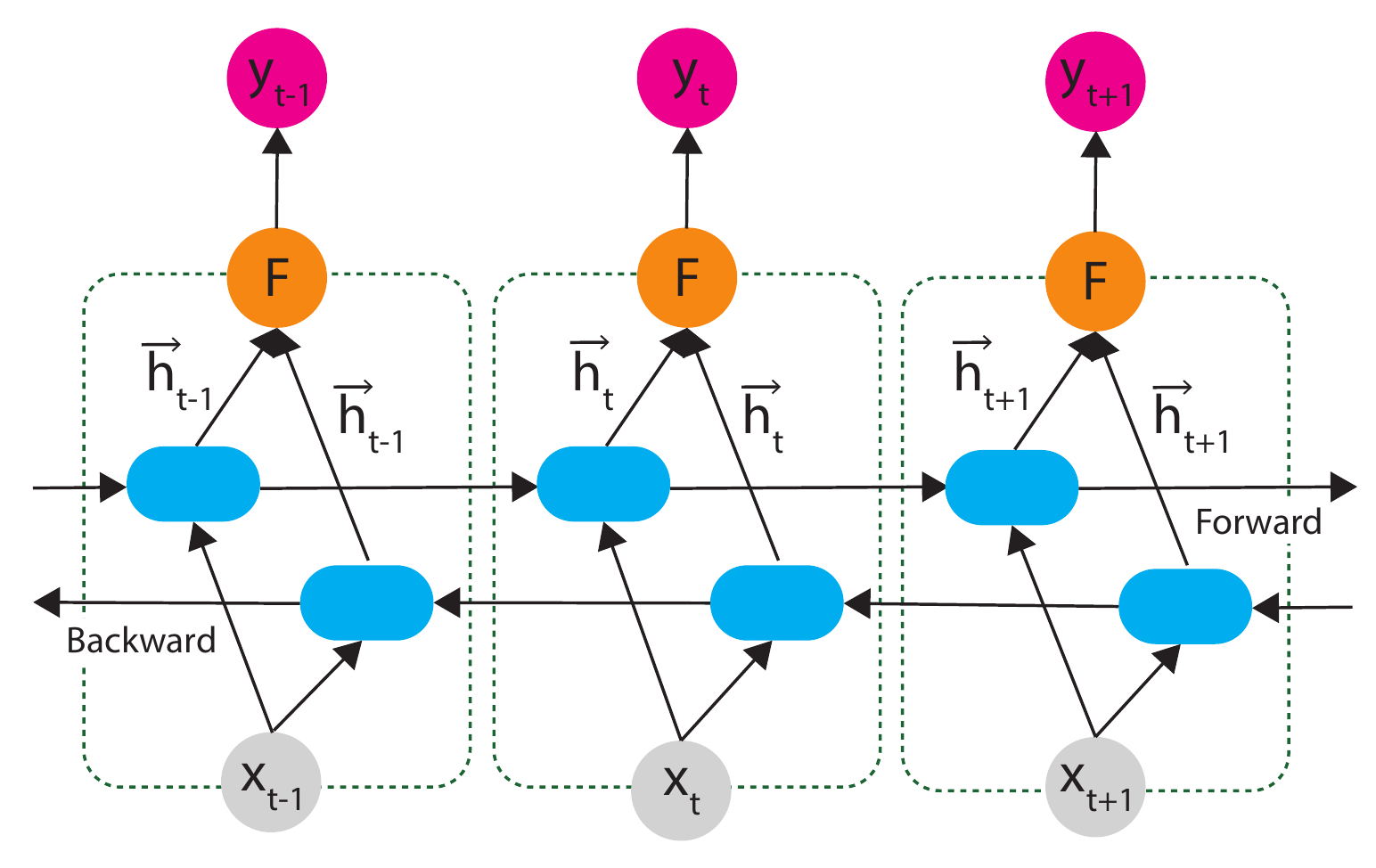}
    \caption{Bidirectional recurrent neural networks. Blue rounded rectangles represent cells and $x_t$
    is the $t$th element of an input sequence $\mathbf{x}$. 
    The input sequence $\mathbf{x}$ is processed by the two disconnected RNNs from beginning to the end and vice versa. 
    Correspondingly, two hidden state vectors forward $\protect\overrightarrow{\mathbf{h}}^{(t)}$ and backward
    $\protect\overleftarrow{\mathbf{h}}^{(t)}$ are updated and passed between the cells of the same layer.}
    \label{fig:fig2}
\end{figure}

In all the RNNs, described above the state at time $T$ captures information 
from the past $\left( x^{(t)}\right),t=1,\ldots,T-1$ as well as the current state $x^{(T)}$ 
to make a prediction. Such RNNs are said to have ``causal'' structure~\cite{goodfellow2016deep} 
and are called unidirectional RNNs.
RNN architectures that combine an RNN that moves forward through time from $t=1$ to $t=T$ with 
another RNN that moves backward through time from  $t=T$ to $t=1$ 
are called bidirectional RNNs.~\cite{graves13,goodfellow2016deep} 
This is realized by duplicating each recurrent layer in the network. 
The two resulting layers have 
separate forward $\overrightarrow{\mathbf{h}}$ and backward $\overleftarrow{\mathbf{h}}$ hidden state vectors.
Forward and backward layers are not
connected to each other.
An input time sequence is provided in the chronological order to the first RNN
layer and is fed in the reversed 
chronological order to the second RNN layer.  Applying RNNs twice
increases the 
amount of input information available to the network and leads to better capturing 
long-term dependencies and, thus, improves the accuracy of the model.~\cite{baldi99}

Bidirectional recurrent NN can be built from the vanilla RNN,
LSTM, and GRU cells resulting in BRNN, BLSTM,~\cite{schuster97} BGRU models,~\cite{lynn19}
respectively. In each case the update equations are similar to Eqs.~\eqref{eq:rnn}-\eqref{eq:gruh}
but applied separately to forward and backward hidden state vectors. 
For example, for BRNN with \textit{tanh} activation function, $\overrightarrow{\mathbf{h}}$ and $\overleftarrow{\mathbf{h}}$,
assuming they have the
identical size, $\overrightarrow{\mathbf{h}}$, $\overleftarrow{\mathbf{h}} \in \mathbb{R}^{k}$, are updated as follows
\begin{eqnarray}
    \overrightarrow{\mathbf{h}}^{(t)} & =& \tanh \left( \mathbf{b}_{\overrightarrow{h}} + \mathbf{W}_{\overrightarrow{h}} \cdot \overrightarrow{\mathbf{h}}^{(t-1)} + \mathbf{W}_{x \overrightarrow{h}} \cdot \mathbf{x}^{(t)}\right),\label{eq:bdrnnf}\\
    \overleftarrow{\mathbf{h}}^{(t)} & =& \tanh \left( \mathbf{b}_{\overleftarrow{h}} + \mathbf{W}_{\overleftarrow{h}} \cdot \overleftarrow{\mathbf{h}}^{(t+1)} + \mathbf{W}_{x \overleftarrow{h}}\cdot \mathbf{x}^{(t)}\right),\label{eq:bdrnnb}
\end{eqnarray}
where $\mathbf{b}_{\overrightarrow{h}}, \mathbf{b}_{\overleftarrow{h}} \in \mathbb{R}^{k}$ are the bias parameters for
the forward and backward hidden states, respectively; 
$\mathbf{W}_{\overrightarrow{h}}, \mathbf{W}_{\overleftarrow{h}} \in \mathbb{R}^{k \times k}$ 
and $\mathbf{W}_{x \overrightarrow{h}}, \mathbf{W}_{x \overleftarrow{h}} \in \mathbb{R}^{k \times m}$ are the
coefficient matrices. In deep BRNN architectures, forward and backward
hidden states of a previous layer are combined
\begin{equation}
   \mathbf{y}^{(t)}=\mathcal{F} \left(\overrightarrow{\mathbf{h}}^{(t)},\overleftarrow{\mathbf{h}}^{(t)}\right), \label{eq:brrnf}
\end{equation}
and passed to the following layer. In Eq.~\eqref{eq:brrnf}  $\mathcal{F}$ is a function that combines 
the two hidden state vectors. It can be a concatenating function
(this work), element-wise addition, multiplication, or averaging.~\cite{keras,tensorflow}
Thus every hidden RNN layer receives and input from both forward 
and backward layers at the preceding layer.

The total number of trainable parameters of the first layer of BRNN is exactly twice the number of trainable parameters in the 
first layer of a unidirectional RNN with the same length of the hidden state vector and same type of RNN cell. 
In a deep bidirectional recurrent NN architecture an $i$th hidden layer with $m$ RNN cells and the length of hidden
vectors $k_i$ outputs to the next RNN layer a tensor $\mathbf{Z} \in \mathbb{R} ^{2k_i \times m}$,
assuming $\mathcal{F}$ is a concatenating function. 
The next $(i+1)$ layer uses $2k_{i+1}(2k_i+k_{i+1}+1)$ trainable parameters, where
$k_{i+1}$ is the length of the hidden state vector of $(i+1)$ layer. Therefore, deep bidirectional recurrent
NNs, in general, use more than twice
the number of trainable parameters than the corresponding unidirectional RNN with the same number of cells
and length of the hidden state vector.

\subsubsection{\label{sec:crnns}Convolutional RNNs}
Convolutional Recurrent Neural networks are deep ANNs that combine
convolutional layers with recurrent layers. In this work, we build two-layer convolutional recurrent NN models 
by adding one recurrent layer (vanilla RNN, GRU, LSTM) to 1D CNN layer resulting in three models denoted as
CRNN, CGRU, and CLSTM. The update equations of each of these types of neural networks are exactly
those of the corresponding individual layers described above. For example, in the CRNN architecture, the first
layer (1D CNN) processes input sequence of length $T$ through convolution, or rather, cross-correlation, operation as given by Eqs.~\eqref{eq:ccs} and~\eqref{eq:ccz}
and generates output tensor $\mathbf{Z}^{(1)}$ formed by the output of $K_1$ kernels (filters)
each of size  $M_1=\lfloor (T-k_1+2p_1)/s_1\rfloor+1$, where $s_1$ is the stride and $p_1$ is the zero padding.
The output tensor $\mathbf{Z}^{(1)}$ is then passed to an RNN layer which is comprised of $M_1$ vanilla RNN 
cells with the hidden state vector of (user-specified) length $k_2$. Each RNN cell receives an input $x^{(t)}\equiv Z^{(1)}_{t,1:K_1}, i\in1,\ldots,M_1$ 
from the preceding 1D CNN layer and processes
it according to the corresponding update equations. 

In this work the recurrent layer of convolutional recurrrent NN models is configured to output the whole sequence of hidden states
from all RNN cells, $\mathbf{Z}^{(2)}\in \mathbb{R}^{M_1\times k_2} $ which is then passed to fully-connected layers.
The total number of trainable parameters of the two-layer architecture described above
is $K_1(k+1)+N_\mathrm{RNN}$,
where $k_1$ is the size of the kernel of 1D CNN layer and $N_\mathrm{RNN}$ the number of trainable
parameters of an RNN layer. 

\subsubsection{\label{sec:cbrnns}Convolutional bidirectional RNNs}
It is possible to fuse a 1D CNN layer and a bidirectional RNN layer. Resulting architectures are called 
Convolutional Bidirectional Recurrent Neural Networks. Such models are expected to combine the benefits of
convolutional recurrent NNs 
with  the improved description of long-term dependencies pertinent to bidirectional 
RNNs. In this work, three convolutional bidirectional NN models are
build by combining one 1D CNN layer with each of the three recurrent NN architectures discussed above: the vanilla RNN, GRU, 
and LSTM. The resulting models are denoted as CBRNN, CBGRU, and CBLSTM. Similarly to convolutional recurrent NNs 
the update equations
of convolutional bidirectional NNs 
can be deduced from the corresponding update equations of 1D CNN and bidirectional recurrent NNs given above. Specifically for
CBRNN, an input sequence is first passed through an 1D CNN layer where it is transformed 
according to Eqs.~\eqref{eq:ccs} and~\eqref{eq:ccz}
into an output tensor $\mathbf{Z}^{(1)}$ which is formed by the output of $K_1$ kernels (filters)
each of size  $M_1=\lfloor (T-k_1+2p_1)/s_1\rfloor+1$, where $s_1$ is the stride and $p_1$ is the zero padding.
Then, $\mathbf{Z}^{(1)}$ is fed separately into forward and backward vanilla RNN layers where the corresponding 
forward and backward hidden states are updated as shown in Eqs.~\eqref{eq:bdrnnf} and~\eqref{eq:bdrnnb}. 
In the two-layer architecture described above, the total number of trainable
parameters is simply the
sum of the total number of trainable parameters of the 1D CNN layer and the total number of trainable
parameters of a bidirectional recurrent layer.


\subsubsection{\label{sec:krr}Kernel ridge regression}
In kernel ridge regression (KRR)~\cite{stulp15,dral19,friedman01}
the approximating function $f(\mathbf{x})$ for a vector of input values $\mathbf{x}$ is defined as
\begin{equation}
    f(\mathbf{x}) = \sum_{i=1}^N \alpha_i k\left(\mathbf{x}, \mathbf{x}_i\right),
\end{equation}
where $N$ is the number of training points and $\boldsymbol{\alpha}=\{\alpha_i\}$ is a vector of regression coefficients.
The covariance function $k\left(\mathbf{x}, \mathbf{x}_i\right)$, commonly referred to as a kernel function or simply kernel, can be understood
as a similarity measure between two vectors $\mathbf{x}$ and $\mathbf{x}_i$ from the input space. The kernel performs
an implicit mapping to a higher-dimensional feature space. 

Because input time-sequences employed in this work have the same length we focus on standard
kernels.~\cite{gneiting10} One of the most common kernel functions is the Mat\'ern kernel~\cite{rasmussen05,gneiting10}, which in the \textsc{MLatom} software package~\cite{dral19mlatom,dral2021mlatom,mlatomdev} used in this work is defined as:~\cite{dral2021mlatom}
\begin{eqnarray}
    k\left(\mathbf{x}_i, \mathbf{x}_j\right)   & = &  \exp \left(-\frac{\| \mathbf{x}_i - \mathbf{x}_j\|_2}{\sigma} \right)
    \sum_{k=0}^n \frac{(n+k)!}{(2n)!}   \nonumber \\
    &\times& \binom{n}{k}  \exp \left(\frac{2\| \mathbf{x}_i - \mathbf{x}_j\|_2}{\sigma} \right)^{n-k}, \label{eq:krrm}
\end{eqnarray}
where $\sigma$
is a positive hyperparameter which defines the characteristic length
scale of the covariance function, $n$ is a non-negative integer, and $\| \ldots\|_2$ is the
Euclidian distance which is taken to be the $L^2$ norm. For $n=0$ the Matern covariance function reduces to
 exponential kernel function~\cite{rasmussen05,gneiting10} 
\begin{equation}
    k\left(\mathbf{x}_i, \mathbf{x}_j\right) = \exp \left( -\frac{\| \mathbf{x}_i - \mathbf{x}_j\|_2}{\sigma}\right). \label{eq:krre}
\end{equation}

 Another popular choice of a covariance 
function is the squared exponential (Gaussian) kernel function~\cite{rasmussen05} 
\begin{equation}
    k\left(\mathbf{x}_i, \mathbf{x}_j\right) = \exp \left( -\frac{\| \mathbf{x}_i - \mathbf{x}_j\|^2_2}{2\sigma^2}\right). \label{eq:krrg}
\end{equation}

Because quantum dynamics often resembles periodically-decaying time-series, we also test a decaying periodic kernel (our adaptation based on Refs.~\citenum{rasmussen05,scikit}):
\begin{equation}
    k\left(\mathbf{x}_i, \mathbf{x}_j\right) = \exp\Bigg(-\frac{\| \mathbf{x}_i - \mathbf{x}_j\|^2_2}{2\sigma^2}  -\frac{2}{\sigma_p^2}\sin^2{\left(\frac{\pi}{p}\| \mathbf{x}_i - \mathbf{x}_j\|_2\right)}\Bigg), \label{eq:krrpd}
\end{equation}
where $p$ is the period and $\sigma_p$ is a length scale for the periodic term (both are hyperparameters).

Given the kernel function, the regression coefficients $\boldsymbol{\alpha}$
are found by minimizing a squared error loss function
\begin{equation}
    \min_{\alpha} \sum_{i=1}^N \left( f\left(\mathbf{x}_i\right) - y_i\right)^2 + \lambda \boldsymbol{\alpha}^T \mathbf{K} \boldsymbol{\alpha}, \label{eq:lsf}
\end{equation}
where $\mathbf{y}=\{y_i\}$ is the target output vector, $\mathbf{K}\in\mathbb{R}^{N\times N}$ 
is the kernel matrix with elements $K_{ij}=k\left(\mathbf{x}_i, \mathbf{x}_j\right)$ 
and $\lambda$ denotes a non-negative
regularization hyperparameter. 
In Eq.~\eqref{eq:lsf}, the second term is usually added to prevent KRR model from assigning large
weight to a single point. The optimization of parameters (regression coefficients $\boldsymbol{\alpha}$) amounts to solving a system of linear equations
\begin{equation}
    \left(\mathbf{K}+\lambda\mathbf{I}\right)=\mathbf{y},\label{eq:krry}
\end{equation}
for which analytical solution is known. Here $\mathbf{I}$ is the identity matrix. The computational scaling of solving this system is
$\mathcal{O}(N^3)$.~\cite{dral2021mlatom,rasmussen05}

KRR is a kernelized version of ridge regression and when linear kernel function
\begin{equation}
    k\left(\mathbf{x}_i, \mathbf{x}_j\right) =\mathbf{x}_i^T\mathbf{x}_j \label{eq:krrl}
\end{equation}
is used, KRR becomes equivalent to ridge regression, i.e., the approximating function $f(\mathbf{x})$ is simply a multiple linear regression with regression coefficients $\boldsymbol{\beta}$ shrunk (see the second term in Eq.~\eqref{eq:lsf}) using regularization:
\begin{eqnarray}
    f(\mathbf{x}) &= &\sum_{i=1}^N \alpha_i \mathbf{x}_i^T\mathbf{x} = \sum_{i=1}^N \alpha_i \sum_{s=1}^{T} x_{is}x_{s} \nonumber \\
    &= &\sum_{s=1}^{T} \left(x_{s}\sum_{i=1}^N \alpha_i x_{is}\right) = \sum_{s=1}^{T} \beta_s x_{s}.
\end{eqnarray}
As shown in above equation, regression coefficients $\boldsymbol{\beta}$ can be conveniently derived from $\boldsymbol{\alpha}$ coefficients and training input data and, in fact, are printed out by \textsc{MLatom}.

\section{\label{sec:comp}Computational details}

\subsection{\label{sec:data}Data sets for training, validation, and testing}
The data set used in this work is the same as used in Ref.~\onlinecite{ullah21} and 
can be accessed at \href{https://doi.org/10.6084/m9.figshare.15134649}{https://doi.org/10.6084/m9.figshare.15134649}.
It was generated as detailed below.~\cite{ullah21} Firstly,
HEOM calculations
for all combinations of the following system and bath
parameters: $\epsilon/\Delta=\{0, 1\}$, $\lambda/\Delta=\{0.1, 0.2, 0.3, 0.4, 0.5, 0.6,0.7, 0.8, 0.9, 1.0\}$, 
$\omega_c/\Delta =\{1, 2, 3, 4, 5, 6, 7, 8, 9, 10\}$, and $\beta \Delta=\{0.1, 0.25, 0.5, 0.75, 1\}$, were performed with QuTiP software package.~\cite{johansson12} In our calculations we set $\Delta = 1.0$.
The total propagation time was $t_\mathrm{max} \Delta=20$ and the HEOM integration time-step was set to $t \Delta=0.05$. 
In total, 1,000 HEOM calculations, 500 for symmetric ($\epsilon/\Delta=0$) and 500 for asymmetric ($\epsilon/\Delta=1$)
spin-boson Hamiltonian, 
were performed. Time-evolved reduced density matrices (RDM) are saved every $dt\Delta=0.1$. 
Secondly, $\langle \hat{\sigma}_z (t)\rangle$ are calculated from RDMs and processed
 into shorter sequences of length $T$ by window slicing.~\cite{rodriguez21,ullah21,lin21}
Namely, for a time series $\mathbf{x}=\left(x^{(1)},\ldots,x^{(L)}\right)$, where
$\langle \hat{\sigma}_z (t)\rangle$ is denoted by $x^{(t)}$ for compactness, a slice is  
a subset of the original time series defined as $\mathbf{s}_{i:j}=\left(x^{(i)},\ldots,x^{(j)}\right), 1\leq i \leq j \leq P$. 
For a given time series $\mathbf{x}$ of length $L$, and the length of the slice is $P$, 
a set of $L-P+1$ sliced time series $\{\mathbf{s}_{1:P},\mathbf{s}_{2:P+1},\ldots,\mathbf{s}_{L-P+1:L}\}$ is 
generated. Finally, the total data set $\mathcal{D}=\{(\mathbf{x}_i,y_i)\}_{i=1}^N$
containing time series $\mathbf{x}_i$ and their corresponding labels $y_i$ is obtained by setting
$1,\ldots,T$ elements of each slice, with $T=P-1$, to an input time-series 
$\mathbf{x}_i$ and the last ($P$th) element of each slice to the associated label $y_i$.

In general, the size of the window $P-1$, or equivalently $T$, should be treated as a hyperparameter
but, following previous work,~\cite{ullah21} we set to $T=0.2L$.
The window slicing is applied to all 1,000 RDMs obtained in HEOM calculations with different system and system-bath parameters. For 
each set of the Hamiltonian parameters the initially calculated set of time-evolved $\langle \hat{\sigma}_z (t)\rangle$ 
with $L=t_\mathrm{max}/dt=200$ generates 160 data points for the data set with $T=41$ (including $t\Delta=0$ point). 

From the raw HEOM data set of 1,000 trajectories, 100 randomly chosen trajectories are taken as the hold-out 
test set, which is used for testing and generating the results presented in Sec.~\ref{sec:results}. 
The remaining set of 900 trajectories are transformed into 144,000 trajectories by window slicing described above.
In total 72,000 short-time $\langle \hat{\sigma}_z (t)\rangle$ trajectories for
symmetric and 72,000 for asymmetric spin-boson models were generated. Each trajectory has
a time-length of  $t\Delta+dt\Delta=4.1$. 
This data set of supervised trajectories is the training set which is randomly partitioned 
into two subsets: a sub-training set, which contains 80\% of the data and a validation set with 20\% 
of the data. ML models are (initially) trained on the sub-training set and the validation set is used for monitoring the performance of the models (mainly, to prevent overfitting and optimize hyperparameters). The final NN models tested on a hold-out test set are not trained on the entire training set and are only trained on the sub-training set to prevent overfitting. KRR models are, however, trained on the entire training set. The performance of KRR models trained only on the sub-training set is similar to the KRR models trained on the entire training set (Appendix~\ref{app:krrsubtrain}). These are typical ways of training NN and KRR models, respectively, which are different due to differences in the formalism and training procedures of these types of models.
Following previous similar works~\cite{rodriguez21,ullah21} the input data is not normalized.

\subsection{\label{sec:anns}Artificial neural network models}
\subsubsection{Details of the models}
In this work fourteen deep ANN models are built and tested. In general, each ANN model is comprised by the total of
two convolutional and/or recurrent layers followed by one fully-connected layer, and one output layer. The exception is the FFNN model
which comprised of two fully-connected layers followed by the output layer. The fully-connected layer in each model, except for FFNN,
has 256 neurons and the rectified linear function defined as (ReLU) $f(z)=\max(0,z)$ is used as the activation function.
 Fixing the properties of
 fully-connected layer allows to compare the performance of recurrent and convolutional layers.
 The output layer contains one neuron with the linear activation function $f(z)=z$. The details of all ANN models
 studied in the present Article is summarized below.

\begin{itemize}
    \item 1D CNN model
contains two 1D CNN layers followed by a MaxPooling, one fully-connected and output layers. 
ReLU is used as the activation function
in each 1D CNN and fully-connected layers. For the MaxPooling layer a pool size $k_p=2$
(see Eq.~\eqref{eq:mp}) is used. The stride of $s=1$ and zero padding are used in both 1D CNN and MaxPooling layers.
The number of filters and the filter sizes of each layer are
optimized using Particle Swarm Optimization algorithm as described in Sec.~\ref{sec:opt}.

\item FFNN model contains two hidden fully-connected layers followed by the output layer. 
The ReLU activation function is used. 

 \item Three recurrent NN models comprised of the two recurrent layers of the same type: 
the vanilla RNN, LSTM, and GRU. The whole sequence of hidden state vectors from all cells is passed from the first  
 to the second recurrent layer as well as from the second recurrent layer to the fully-connected layer.
 
 \item Three bidirectional recurrent NN models comprised of the two bidirectional recurrent layers of the same type: 
 the vanilla BRNN (denoted simply as BRNN hereafter), 
 BLSTM, and BGRU. The whole sequence of hidden state vectors from all cells is passed from the first bidirectional recurrent layer 
 to the second bidirectional recurrent layer as well as from the second bidirectional recurrent layer to the fully-connected layer.
 
 \item Three convolutional recurrent NN models comprised of one 1D CNN layer and one recurrent layer of each type: 
 the vanilla RNN, LSTM, and GRU. The resulting models are denoted as CRNN, CLSTM, and CGRU.
 The recurrent layer returns the whole sequence of hidden state vectors from all cells to the fully-connected layer.
 The stride of $s=1$, zero padding, and the ReLU activation function are used in the 1D CNN layer. The number of filters and the filter size is the same as in the first layer of the 1D CNN model.

 \item Three convolutional bidirectioinal recurrent NN models comprised of one 1D CNN layer and one bidirectional
 recurrent layer of each type: 
 the vanilla RNN, LSTM, and GRU. The resulting models are denoted as CBRNN, CBLSTM, and CBGRU.
 The stride of $s=1$, zero padding, and the ReLU activation function is used in the 1D CNN layer. The number of filters and the filter size is the same as in the first layer of the 1D CNN model. The bidirectional recurrent layer returns the whole sequence of the forward-backward hidden state vectors from all cells to the fully-connected layer.
\end{itemize}

\subsubsection{\label{sec:opt} Hyperparameter optimization}
The goal of this work is to compare the performance of ML models including ANN models with different architectures as described above. 
Depending on the overall objective, one can envision several approaches for 
comparing performances of different ANNs. 
For example, the hyperparameters of each of ANN model can be adjusted using grid search, random search, or other optimization
techniques, including evolutionary algorithms,
to achieve the best performance of each ANN model on the given data set. However, given the
fundamentally different types of ANN models considered in the present Article, this approach may result in
several models with approximately the same prediction accuracy but requiring drastically different 
resources such as CPU (or GPU) time and memory. In such case, it might be reasonable to use ANN models
that are not the most accurate, but provide an acceptable accuracy with less training and prediction times. 
The notion of the optimal
running time, however, strongly depends on the problem under study. One can also set the desired accuracy
level and systematically adjust the hyperparameters of each model to achieve the desired accuracy
and then analyze the resulting ANN models. This approach, however, requires specifying the fixed accuracy level
which might not be achievable for some ML models and relies on \textit{a priori} knowledge of the dynamics
of the system of interest.

The following approach is adopted in the present study. We compare the performance of ANN models with
approximately the same number of trainable parameters. Hyperparameters of each ANN model described above
are scanned using a grid search approach and the models with the pre-determined number of trainable parameters
are selected and their performance is reported in Sec.~\ref{sec:results}. The number of trainable parameters
is chosen as follows. In Refs.~\citenum{rodriguez21} and~\citenum{ullah22} we illustrated that deep 1D CNN models can approximate long-time dynamics of a molecular dimer system (spin-boson-like model) and the
Fenna--Matthews--Olson photosynthetic complex accurately. Therefore, 
the 1D CNN model from Ref.~\citenum{rodriguez21} is taken as a base model. The hyperparameters of this model, associated with 
convolutional layers, specifically,
the number and the size of the kernels of each layer,
are optimized using Particle Swarm Optimization algorithm (PSO). All other hyperparameters such as the number of
neurons of a fully-connected layer and the activation functions of all neurons are fixed to 256 and ReLU correspondingly.
PSO calculation is performed with three particles and for 50 steps. The data set described above is used for the
hyperparameter optimization. 
At each step of the PSO calculation,
three 1D CNN models, one for each particle, are trained and validated using the above mentioned data set.
During training, the deviation between predicted $\hat{y}_i$
and reference values $y_i$ 
is minimized. In this work, we use the mean squared
error (MSE) as the loss function in the minimization
\begin{equation}
    \mathrm{MSE} = \frac{1}{N_b}\sum_{i=1}^{N_b} \left(y_i-\hat{y}_i\right)^2.
\end{equation}
Adaptive moment estimation (Adam) algorithm~\cite{kingma17} is used with the initial learning rate set to 1.0$\cdot$10$^{-4}$.
The initial values of the weights 
are randomly sampled using Xavier initialization.~\cite{glorot10} The biases are initialized to zero. 
\textsc{Keras}~\cite{keras} 
software package with the \textsc{TensorFlow}~\cite{tensorflow} backend was employed for calculation. The batch size is set 
to $N_b=64$ and the training for each model is performed for 30 epochs. As will be shown later this is
sufficient for the MSE to drop below 10$^{-5}$--10$^{-6}$.

For each trained 1D CNN model, the MSE for the validation set is calculated and used as the fitness value
in the PSO. Then the hyperparameters of 1D CNN models are adjusted based on the algorithmic details of PSO
which can be found in Appendix~\ref{app:pso}.
After 50 steps of the PSO, the 1D CNN models stop improving and the PSO calculation is terminated.
The optimized number of kernels of the final 1D CNN model is 235 and 125 
with the kernel sizes 16 and 7 of the first and second layer, respectively.
The 1D CNN model described above contains a total of 530,258 trainable parameters. This number is taken as 
reference and the hyperaprameters of all other 13 ANN models are selected to generate models
with approximately the same number of parameters. Note, however, that according to the formulas 
given in Sec.~\ref{sec:intro}, it is not possible to set hyperparameters in all models to
achieve exactly the desired number of trainable parameters. Therefore, some variation in the number
of trainable parameters across the reported ANN models is to be expected.

Even though the number of trainable parameters is straightforwardly connected to the hyperparameters
of each ANN architecture, a grid search is performed over hyperparameters with the goal to
understand the sensitivity of the model performance to the small variations in the hyperparameters. In all
models containing (bidirectional) recurrent layers the number of units is scanned from 5 to 100 with the step of 5.
Additionally, in all ANN models containing 1D CNN layers and (bidirectional) recurrent layers, the number of kernels and kernel sizes
of the 1D CNN layer is taken to be same as in 1D CNN model while the number of units in the recurrent layer is scanned
from 5 to 100 in steps of 5. 
All the models are trained, validated, and tested using the
 data set for the symmetric spin-boson model. \textsc{keras} software package and Adam algorithm is used for all ANN calculations.
The learning rate is fixed to 1.0$\cdot$10$^{-4}$, the batch size is set to 128, and 
the training for each model is performed for 30 epochs. 

Usually more than one set of hyperparameters generates ANN
models with the
number of trainable parameters close to the target number. In such cases, models with the number of trainable parameters within $\pm$5\% of the
reference number are considered.  Among selected models, the models that perform significantly
different than the average model are discarded. The number of outliers in each case is found to be small.
We attribute outliers to finite training time and nonuniform decay of the MSE during the training, which in case of 
RNNs is a manifestation of the vanishing gradient problem. We will
return to this issue in Sec.~\ref{sec:results}.
The model with the closest number of trainable parameters to the target number
is saved and its performance is reported.

\subsubsection{Training and validation curves}
\begin{figure*}
    \centering
    \includegraphics[width=0.99\textwidth]{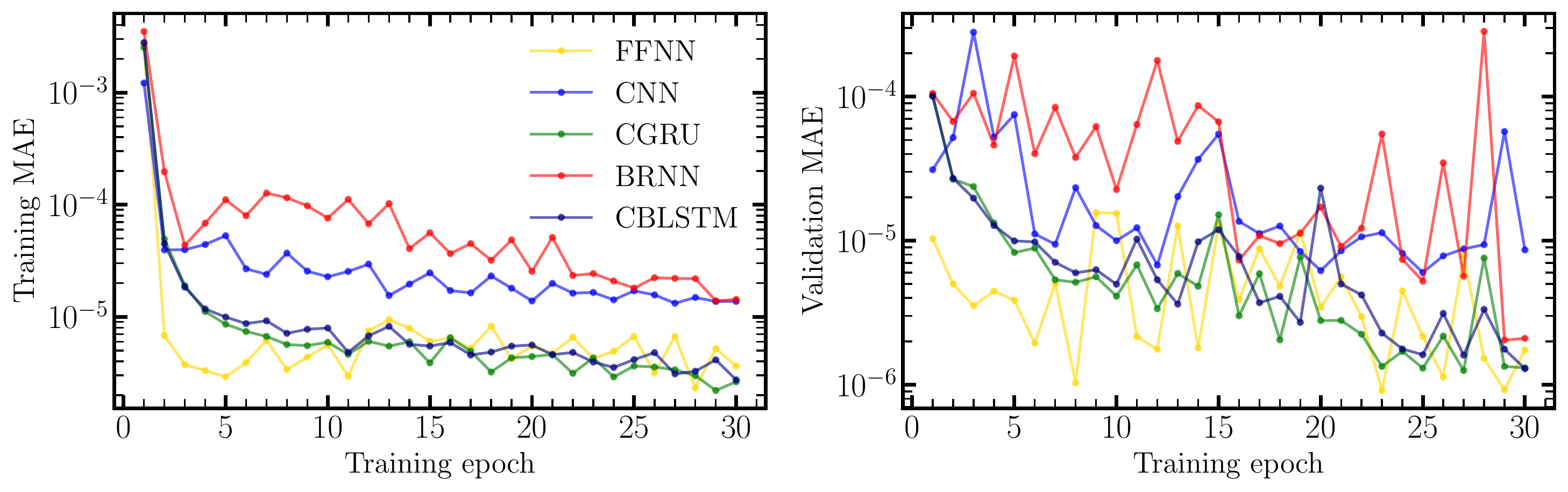}
    \caption{Learning curves for selected ANN models: training (left) and validation (right). Note the logarithmic scale of the vertical axis.}
    \label{fig:tr}
\end{figure*}

To illustrate the learning process of ANN models in Fig.~\ref{fig:tr} we plot the training and validation curves for several such models.
Such curves are a widely used tool to examine the performance of supervised learning algorithms. In general
the learning 
is found to be stable in each case. The apparent noisiness in the data
should be attributed to the logarithmic scale on the vertical axis. Several conclusions can be
drawn by examining the curves shown in Fig.~\ref{fig:tr}. Firstly, none of the models overfits which is illustrated in lower
and in, general, decaying validation MAE (Fig.~\ref{fig:tr} right panel) compared to the training MAEs (Fig.~\ref{fig:tr} left panel).
Secondly, the learning process is fast even with the learning rate of 1$\cdot 10^{-4}$.
It should be noted that decreasing learning rate to 1$\cdot 10^{-5}$ does not result in noticeable improvement in accuracy for
all the ANN models studied in this work
but the increase of the learning rate to 1$\cdot 10^{-3}$ usually results in increasing MAE.
Furthermore, FFNN model reaches the lowest MAE rapidly in less than 10 training epochs while the MAE of other models e.g., 1D CNN, B(LSTM,GRU) 
continues to decay over all shown 30 training epochs. The same behavior is manifested in the validation MAEs.
This suggests that 30 training epochs used in this work is justified for 1D CNN and other convolutional and recurrent models
but seems unnecessarily too many for the FFNN model.
This observation further illustrates the efficiency of simple FFNN models. 
We note that in the case of the BRNN model the validation MAE exhibits
large-magnitude changes over the course of training. This can be attributed to the vanishing gradient problem of the vanilla
RNNs.

\subsection{\label{sec:krrs}KRR models}
In addition to 14 ANN models, 8 KRR models are investigated in the present Article. Each KRR 
model differs by the choice of the kernel.
The following kernels are studied: 
the linear kernel denoted as KRR-L (Eq.~\eqref{eq:krrl}),  Gaussian kernel (Eq.~\eqref{eq:krrg}, KRR-G), 
exponential kernel Eq.~\eqref{eq:krre} (KRR-E), Mat\'ern kernel, Eq.~\eqref{eq:krrm} with $n=1,2,3,4$ (KRR-M1, KRR-M2, KRR-M3, KRR-M4),
and a periodic-decaying kernel, Eq.~\eqref{eq:krrpd} (KRR-DP). 
It should be noted
that unlike recurrent neural networks, KRR (and FFNN) models studied in this work require a fixed-size input.
To enable the comparison between ANN and KRR models studied in 
this work, we focused on the input $\langle \hat{\sigma}_z(t)\rangle$ trajectories of the same length.
We note that extending RNN models built in the present Article to variable size input is straightforward
and will be discussed elsewhere.

In KRR models, optimization amounts to finding regression coefficients $\boldsymbol{\alpha}$
as shown in Eq.~\eqref{eq:krry}. Therefore, one can interpret each regression
coefficient $\alpha_i$ as a trainable parameter. The total number of such coefficients
is the same as the number of elements of the training set, which in this work is 72,000
in each case of symmetric and asymmetric spin-boson models.
This number of trainable parameters of KRR model is 
far fewer than the target number of trainable parameters for ANN models.
The discrepancy between the number of trainable parameters which, by construction, is set
to the size of the training set, yet it is variable in ANN models. This further makes the faithful comparison
of fundamentally different ML approaches studied in this work non-trivial. As will be shown
in the following, the fewer number of trainable parameters of KRR does not adversely
impact the performance of KRR models.

KRR approaches have analytical solution to optimal parameters (regression coefficients $\boldsymbol{\alpha}$), but they still require adjusting a (small) number of
hyperparameters. All KRR models used here contain the regularization hyperparameter
$\lambda$ which is the only hyperparameter in KRR-L. Other KRR models have additional hyperparameters in their kernel functions, i.e., all nonlinear models have length-scale hyperparameter $\sigma$, KRR-M$n$ also have the integer hyperparameter $n$ (not optimized here), KRR-DP has two more hyperparameters compared to KRR-G: $p$ and $\sigma_p$. Due to the small number of hyperparameters, they are optimized on a logarithmic grid as described elsewhere\cite{dral19mlatom} for all models except for KRR-DP which has three hyperparameters. Hyperparameters in KRR-DP are optimized using the tree-structured Parzen estimator~\cite{tpe} via interface to the \textsc{hyperopt} package~\cite{Bergstra2015HyperoptAP}. All hyperparameters of KRR models are
optimized based on the data set for symmetric spin-boson model 
using grid-search method for KRR.

The hyperparameters of each ML model are optimized only for the data set for symmetric spin-boson model
as already indicated above. The optimized models are then re-trained on the asymmetric spin-boson data set
keeping the hyperparameters unchanged. \textsc{MLatom} software package~\cite{dral19mlatom,dral2021mlatom} is used for all KRR calculations.

\section{\label{sec:results}Results and discussion}
\subsection{Symmetric spin-boson model}

The details and performance of all ANN and KRR models are summarized in Table~\ref{tab:1} and Table~\ref{tab:2}, respectively.  As seen from Table~\ref{tab:1}, the two best performing ANN models on the symmetric
spin-boson data set, are the convolutional LSTM (CLSTM) and convolutional bidirectional LSTM (CBLSTM) while the two least accurate ANN models are
unidirectional RNN and bidirectional vanilla RNN models. Worse performance of 
simple vanilla RNN layers compared to LSTM 
and GRU layers is the expected result. It has been observed in other applications that in contrast to LSTM and GRU models,
the performance of vanilla RNN models degrades with the increasing length of input sequence.~\cite{sterin17,gupta20} 
We note that the weak performance
of RNN models is despite the increased number of units or length of hidden state vector compared to other recurrent NN models.

\begin{table*}
\caption{Hyperparameters, mean absolute prediction errors, total number of 
trainable parameters, training, and average single step prediction 
times of all ANN models studied in this work. For each recurrent layer the number of
units is shown. The number of kernels X and kernel sizes Y for each convolutional layer is shown in
parenthesis as (X,Y).}  
\begin{ruledtabular}
\begin{tabular}{l*{7}c} 
       & \multicolumn{1}{c}{Trainable} & \multicolumn{2}{c}{Layers} 
       & \multicolumn{2}{c}{Mean absolute error} & \multicolumn{2}{c}{Time [s]} \\  \cline{3-4} \cline{5-6} \cline{7-8}
 Model & \multicolumn{1}{c}{Parameters} & \multicolumn{1}{c}{Layer 1} & \multicolumn{1}{c}{Layer 2}  &   \multicolumn{1}{c}{Symmetric} & \multicolumn{1}{c}{Asymmetric} & \multicolumn{1}{c}{Training} & \multicolumn{1}{c}{Prediction}\\ \hline
 1D CNN & 530,258 & (235,16) & (125,7) &  1.55$\cdot 10^{-3}$   & 4.84$\cdot 10^{-2}$ & 465 & 3.8\\
 FFNN     & 520,045 & 754 & 646 & 1.32$\cdot 10^{-3}$ & 3.70$\cdot 10^{-2}$ & 82 & 3.3\\
 \multicolumn{7}{c}{Recurrent Neural Networks}\\
 LSTM   & 528,577 & 15 & 49 &   1.58$\cdot 10^{-3}$  & 2.35$\cdot 10^{-2}$ & 623 & 6.1\\
 GRU    & 553,453 & 60 & 50 &  2.05$\cdot 10^{-3}$  & 2.57$\cdot 10^{-2}$ & 668 & 4.4\\
 RNN    & 535,468 & 65 & 50 &  3.00$\cdot 10^{-3}$  & 6.17$\cdot 10^{-2}$ & 302 & 4.2\\
 \multicolumn{7}{c}{Convolutional Recurrent Neural Networks}\\
 CLSTM  & 501,965 & (28,16) & 71 &  1.17$\cdot 10^{-3}$   & 2.50$\cdot 10^{-2}$ & 279 & 4.1\\
 CGRU  & 515,806 & (55,16) & 73 &   1.38$\cdot 10^{-3}$  & 2.14$\cdot 10^{-2}$ & 294 & 4.7\\
 CRNN  & 513,673 & (243,16) & 73 &   1.46$\cdot 10^{-3}$  & 3.61$\cdot 10^{-2}$ & 197 & 5.4\\
 \multicolumn{7}{c}{Convolutional Bidirectional Recurrent Neural Networks}\\
 CBLSTM  & 568,022 & (109,16) & 39 &  1.17$\cdot 10^{-3}$    & 2.84$\cdot 10^{-2}$ & 333 & 3.8\\
 CBGRU  & 514,860 & (55,16) & 37 &  1.54$\cdot 10^{-3}$   & 3.68$\cdot 10^{-2}$ & 325 & 5.3\\
 CBRNN  & 508,842 & (297,16) & 36 &   2.50$\cdot 10^{-3}$  & 3.58$\cdot 10^{-2}$ & 256 & 3.9\\
 \multicolumn{7}{c}{Bidirectional Recurrent Neural Networks}\\
 BLSTM  & 511,809 & 6 & 24 &  2.12$\cdot 10^{-3}$   & 2.56$\cdot 10^{-2}$ & 635 & 5.2\\
 BGRU  & 534,991 & 14 & 25 &   2.28$\cdot 10^{-3}$  & 2.48$\cdot 10^{-2}$ & 1109 & 6.7\\
 BRNN  & 511,959 & 37 & 24 &   6.95$\cdot 10^{-3}$  & 4.27$\cdot 10^{-1}$ & 396 & 5.4
\end{tabular}
\end{ruledtabular}
\label{tab:1}
\end{table*}

A better performance
of convolutional recurrent NN models 
is an illustration of the power of sub-sampling of
the initial input data which is then more efficiently learned by the LSTM layer. 
Comparing the unidirectional or bidirectional 
recurrent NN models whether taken separately or as the second layer on top of 1D CNN layer confirms the trend. The best performing
model is an (C,B,CB)LSTM model, followed by the (C,B,CB)GRU, and (C,B,CB)RNN. It is also worth noting that recurrent NN models
based on bidirectional layers do not outperform their unidirectional counterparts. This is observed for both unidirectional recurrent 
models compared to bidirectional recurrent models as well as when convolutional unidirectional recurrent models are compared to
the corresponding convolutional bidirectional recurrent models. 

\begin{table*}
\caption{Mean absolute prediction errors (MAEs), training, and 
average single step prediction times of all KRR models used in this work. 
KRR-L, KRR-G, KRR-DP, KRR-E, and KRR-M ($n=1,2,3,4$) denote  kernel ridge regression models with linear kernel, 
Gaussian kernel, decaying-periodic kernel, exponential kernel, 
and Matern kernel with $n = 1,2,3,4$, respectively.}  
\begin{ruledtabular}
\begin{tabular}{l*{6}c} 
       & \multicolumn{1}{c}{Trainable} 
       & \multicolumn{2}{c}{Mean absolute error} & \multicolumn{2}{c}{Time [s]} \\  \cline{3-4} \cline{5-6}
 Model & \multicolumn{1}{c}{Parameters}  &  \multicolumn{1}{c}{Symmetric} & \multicolumn{1}{c}{Asymmetric} & \multicolumn{1}{c}{Training} & \multicolumn{1}{c}{Prediction} \\ \hline
KRR-L &  72,000  & 1.2$\cdot10^{-2}$  & 6.5$\cdot10^{-2}$ & 196 & 1.6\\
 KRR-G &  72,000  & 4.7$\cdot10^{-4}$  & 1.2$\cdot10^{-3}$  & 220 & 1.6\\
 KRR-DP &  72,000  & 4.3$\cdot10^{-4}$  & 2.0$\cdot10^{-3}$ &  257 & 1.4\\
 KRR-E &  72,000  & 2.1$\cdot10^{-3}$  & 3.3$\cdot10^{-3}$ &   222 & 1.6\\
KRR-M1 &  72,000   & 2.4$\cdot10^{-4}$  & 1.3$\cdot10^{-3}$ & 260 & 1.6 \\
KRR-M2 &  72,000  & 2.2$\cdot10^{-4}$ & 2.7$\cdot10^{-3}$ & 259 & 1.7 \\
KRR-M3 &  72,000  & 2.0$\cdot10^{-4}$ & 2.3$\cdot10^{-3}$ & 279 & 1.7 \\
KRR-M4 &  72,000   & 2.3$\cdot10^{-4}$ & 2.1$\cdot10^{-3}$ & 273 & 1.7 \\
\end{tabular}
\end{ruledtabular}
\label{tab:2}
\end{table*}

Focusing on KRR models, whose details and performance is illustrated in Table~\ref{tab:2}, we note that these
 models generally outperform ANN models with the exception of a KRR model with linear kernel. Furthermore, the
 KRR-L is the worst
performing  of all ML models studied in the present work. This is, perhaps, not surprising because it is expected
that quantum dynamics of such a complex system as the spin-boson model
is highly non-trivial and cannot be captured by
simple linear regression.

\begin{figure*}
    \centering
    \includegraphics[width=0.99\textwidth]{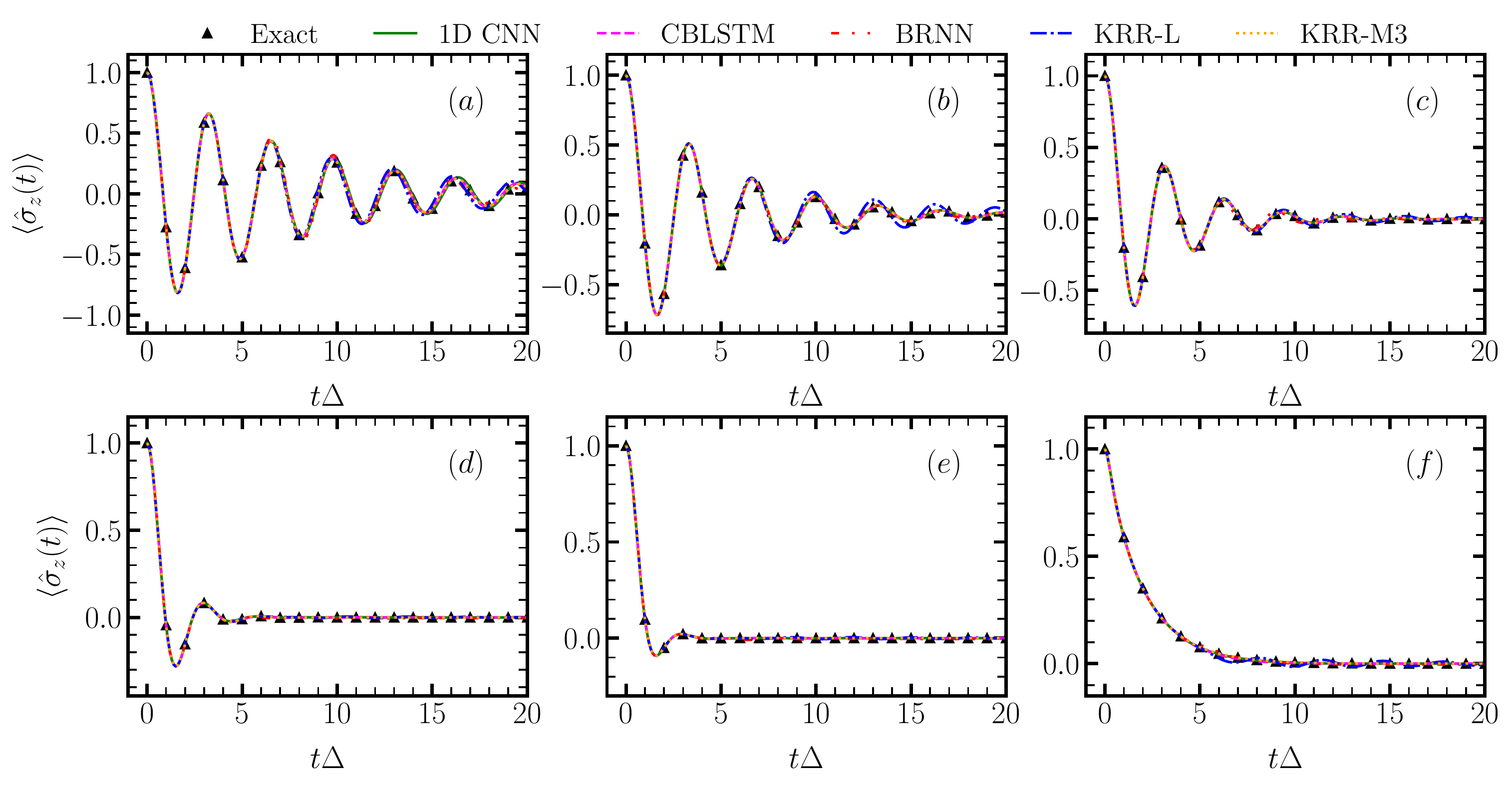}
    \caption{Expectation values $\langle \hat{\sigma}_z(t)\rangle$ for symmetric spin-boson model as a function of time.
    Results predicted by an indicated ML model are compared to the HEOM results (black triangles) for the following parameters:
    (a) $\lambda=0.2, \omega_c=8.0, \beta=1.0$; (b) $\lambda = 0.4, \omega_c=10.0,
    \beta=1.0$; (c) $\lambda=0.2, \omega_c=10.0, \beta=0.25$; (d) $\lambda=0.1, \omega_c=4.0, \beta=0.1$; 
    (e) $\lambda=0.8, \omega_c=3.0, \beta=1.0$; (f) $\lambda=1.0, \omega_c=2.0, \beta=0.1$. 
    All parameters are in the units of $\Delta$.}
    \label{fig:sym}
\end{figure*}

The differences in accuracy between ML models noted above is, however, not significant in the case of symmetric spin-boson model.
This is illustrated in Fig.~\ref{fig:sym}
where the exact population difference $\langle \hat{\sigma}_z(t)\rangle$ is compared to the ML-predicted $\langle \hat{\sigma}_z(t)\rangle$.
We stress that only short $\langle \hat{\sigma}_z(t)\rangle$ trajectory of $t\Delta=4$
is used as an input. The rest of the dynamics is predicted 
recursively as done in Refs.~\citenum{rodriguez21} and~\citenum{ullah21}.
Fig.~\ref{fig:sym} compares the performance of the 1D CNN model, whose number of trainable
parameters is used as a reference for other ANN models, to the best and the worst
performing ANN and KRR models. One notices
that prediction accuracy of the KRR-L model degrades slowly over time but still remains acceptable. This can only be
seen in Fig.~\ref{fig:sym}a and b where the chosen system and bath 
parameters generate the oscillatory
dynamics of RDM. In the cases of incoherent relaxation dynamics
KRR-L results are indistinguishable from the exact HEOM dynamics.  

We stress that even the worst performing ML model, KRR-L, as illustrated in
Fig.~\ref{fig:sym} still provides the acceptable accuracy. 
Therefore, we conclude that all ML models benchmarked in the present article can be trained
to provide a sufficient long-time prediction accuracy. The suspected drawback of the recursive propagation approach proposed in 
Refs.~\citenum{rodriguez21} and~\citenum{ullah21} and exploited in this article is that the prediction error would grow over many time
steps possibly leading to significant accuracy loss in the long-time dynamics. As shown above such errors can be made
small enough to make long-time predictions reliable all the way until the dynamics reaches equilibrium. 

\subsection{Asymmetric spin-boson model}
The results reported so far are encouraging but not discriminative.
To unravel the differences between the studied ML models, we devise a more stringent test. All the ANN and KRR models  are re-trained on the
asymmetric spin-boson model data set without any  hyperparameter adjustment. 
The increased difficulty of this test stems from the richer dynamics of the asymmetric spin-boson model
which might require more training parameters or even different ANN architecture (more hidden layers,
longer hidden state vectors, and memory $T$).

The proposed test clearly indicates that some models predict the long-time dynamics much more accurately than others.
The performance of each model on the hold-out test set is reported in Tables~\ref{tab:1} and~\ref{tab:2}. There is a noticeable
increase in the MAE of all studied ML models compared to the symmetric spin-boson model. Generally, KRR models are  more
robust than ANN models. On average, the MAE of KRR models increases by a factor of 6. KRR models based on M\'atern kernels with $n=2,3,4$ exhibit the most
significant performance drop of $\sim$10 times. The average decrease in performance of ANN models is $\sim$18.
Interestingly, the most significant
performance reduction, among the ANN models, is observed for the models that do not contain recurrent layers: 1D CNN and FFNN while models
containing bidirectional recurrent layers revealed to be the most robust ANN models. On average the MAE of 
B(LSTM, GRU, RNN) models decreased by a factor of $\sim$10 while, for example, the FFNN model became $\sim$30 times less
accurate for the assymetric spin-boson model.

\begin{figure*}
    \centering
    \includegraphics[width=0.99\textwidth]{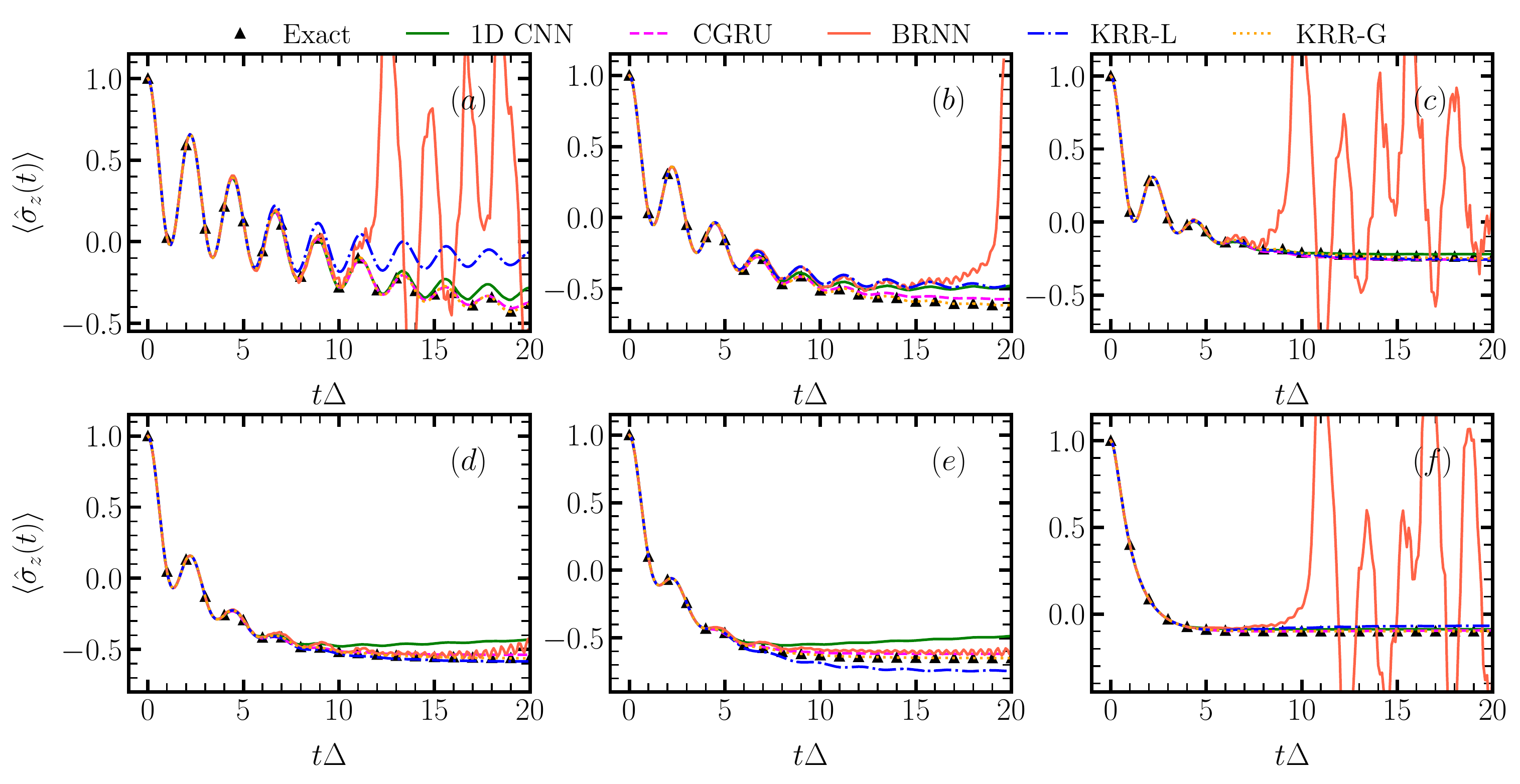}
    \caption{Expectation values $\langle \hat{\sigma}_z(t)\rangle$ for asymmetric spin-boson model with
    $\epsilon=1$ as a function of time.
    Results predicted by an indicated ML model are compared to the HEOM results (black triangles) for the following parameters:
    (a) $\lambda=0.1, \omega_c=6.0, \beta=0.75$; (b) $\lambda = 0.3, \omega_c=8.0,
    \beta=1.0$; (c) $\lambda=0.2, \omega_c=10.0, \beta=0.25$; (d) $\lambda=0.4, \omega_c=8.0, \beta=0.75$; 
    (e) $\lambda=0.8, \omega_c=10.0, \beta=1.0$; (f) $\lambda=0.7, \omega_c=10.0, \beta=0.1$. 
    All parameters are in the units of $\Delta$.}
    \label{fig:asym}
\end{figure*}

Fig.~\ref{fig:asym}  shows the ML-predicted dynamics of $\langle \hat{\sigma}_z(t)\rangle$ of the most and least
accurate ANN and KRR models for the asymmetric spin-boson model as well as of the ``reference'' ANN model, 1D CNN. 
Analogously to the symmetric spin-boson model,
BRNN model is found to be least accurate ANN model. The difference is, however, more dramatic. Clearly, BRNN model
is overall no longer acceptable ANN model even though it provides an accurate prediction for some parameters, 
see Fig.~\ref{fig:asym}d and e, but even in these cases one can easily distinguish an unphysical oscillatory behavoir
building up beyond relatively short times of $t\Delta \approx 10$. 
All other ANN models provide more accurate and stable
predictions without noise and unphysical artifacts. The CGRU model is the most accurate matching the HEOM dynamics nearly exactly.
The worst KRR model is, once again, the one with the linear kernel. It is however, able to capture the dynamics
qualitatively, and in some, cases fairly accurately. It seems to be less reliable in predicting the oscillatory
dynamics typically observed in low temperature and small reorganization energy regimes. The most accurate KRR model is
the one with the Gaussian kernel. This model predicts the
reference HEOM dynamics very accurately. The promise of the KRR-G model has been already pointed out in Ref.~\citenum{ullah21}.

\subsection{Training and prediction times}
The accuracy and robustness of ML models are clearly very important but our analysis would be incomplete without discussing the timings
associated with all studied ML models. In large-scale applications,
the choice of ML model is often a compromise between the accuracy and the complexity of the model. The latter directly affects 
the training as well as the prediction times which are important, especially for the approaches 
based on recursive application of ML
models to generate the dynamics. Training times for all ML models are calculated using 2x20C Intel Xeon Gold 6230 2.1GHz processor
with 192 Gb DDR4 memory, 
and reported in Tables~\ref{tab:1} and~\ref{tab:2} (only times for training are shown, in practice, the cost is higher if the hyperparameter optimization is performed). One should be mindful that different software
packages are used for ANN and KRR models which makes the direct comparison somewhat dubious.
Nonetheless both \textsc{keras} and \textsc{MLatom} are state-of-the-art software widely used in the community
so we will proceed with the comparison.

According to Table~\ref{tab:1}, the accuracy of FFNN and CBGRU models for the asymmetric
spin-boson model is the same but the training time of the FFNN model is nearly 4 times shorter.
Furthermore, for the symmetric spin-boson Hamiltonian,
FFNN model outperforms all ANN models comprised of only recurrent layers but shows similar performance to convolutional
recurrent neural networks and bidirectional recurrent neural networks and it does so faster: FFNN model is 3 times faster to train
than the best performing CLSTM and CGRU models. 

Continuing the comparison of the training times we note that, due to their simplicity, the vanilla  (unidirectional) RNN models
are also fast to train. Oppositely, bidirectional RNNs of all kinds are among the slowest models to train. Given that
using bidirectional architecture  does not reduce the error compared to their unidirectional counterparts
and the increased training time of bidirectional layers makes such models disfavoured for, at least, the problem of 
predicting the long-time dynamics of the spin-boson problem. This result is interesting given that bidirectional recurrent neural 
networks have some success in other domains as described in Introduction.

Training times of the KRR models is relatively fast compared to most of the NN models except for FFNN and CRNN.

The average single-step prediction times for each ANN and KRR models are shown in the last column of
Tables~\ref{tab:1} and ~\ref{tab:2}. 
It is clear that KRR models take less time to make a single time-step prediction than ANN models, although this can
be attributed to the software nuances. 
Expectedly, the FFNN model is the fastest ANN model while BGRU is the slowest. Similar to the training time we observe
that (C,B)GRU models are slower than (C,B)LSTM models which is counerintuitive given than GRU cell is less complex
than the LSTM one. We attribute this inconsistency to the implementation details of both cells in \textsc{keras}.
Overall, recurrent layers require more time to evaluate than other layers as should be anticipated. 
Bidirectional recurrent neural networks take even more
time to make a prediction compared to their unidirectional
counterparts. Convolutional recurrent neural network models take less time to predict compared
to their two-layer recurrent counterparts as they replace one computationally more expensive 
recurrent layer with a chaper 1D CNN layer.

\begin{figure*}
    \centering
    \includegraphics[width=0.98\textwidth]{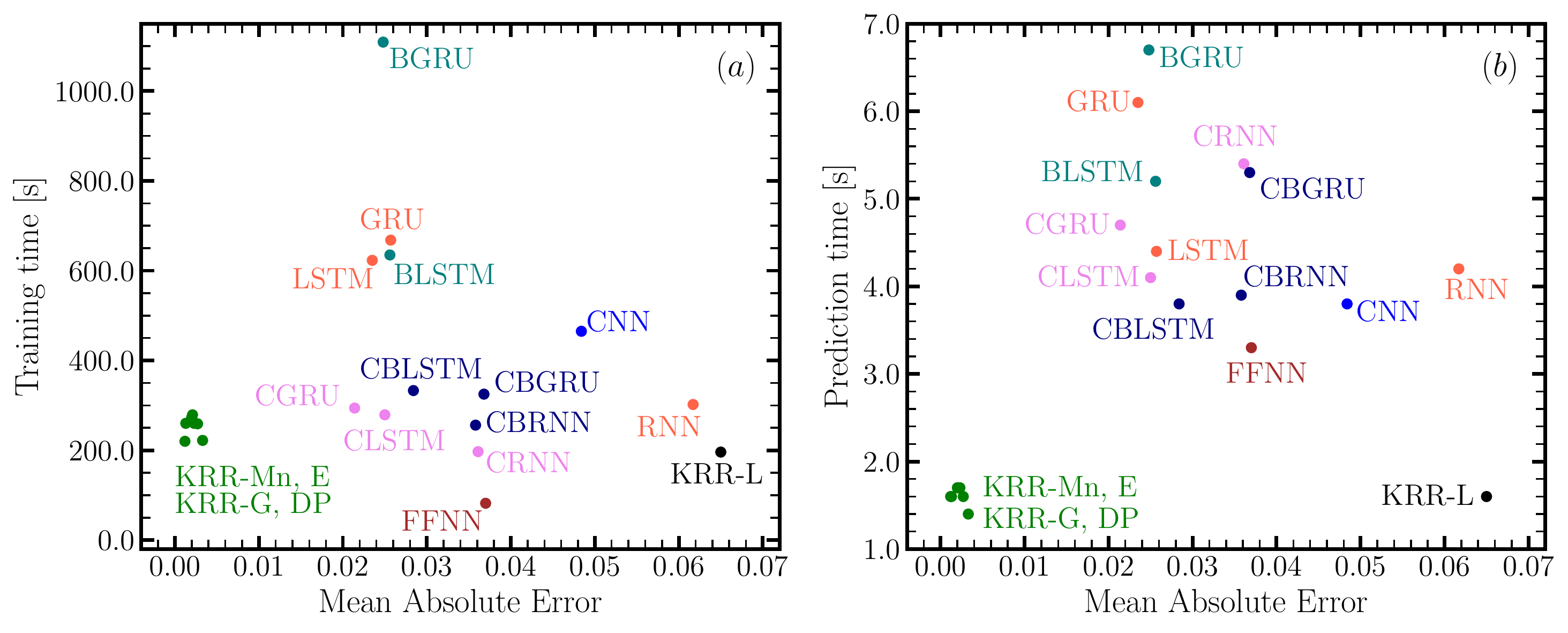}
    \caption{The mean absolute single-step prediction error of $\langle \hat{\sigma}_z(t)\rangle$ 
    of machine learning models studied in this work 
    plotted against the corresponding training \textit{(a)} and single-step
    prediction \textit{(b)} times. 
    Bidirectional recurrent neural network model is not shown because its 
    error is much too big compared to other models (see Table~\ref{tab:1} and Fig.~\ref{fig:asym}).}
    \label{fig:all}
\end{figure*}

\subsection{Model efficiency}
To illustrate the tradeoff between accuracy and training/prediction time, 
in Fig.~\ref{fig:all} the training and single-step prediction times are 
plotted against the MAE for each ML model built 
in the present work. For ease of illustration, the models of similar kind are highlighed
with the same color, e.g., all KRR models are depicted by green circles, magenta color is used for all convolutional recurrent models, etc. 
Clearly, KRR models are the most efficient models showing high accuracy and requiring the shortest amount of 
time to train and make a prediction. 
Among the ANN models, convolutional recurrent models are the most efficient.
According to Fig.~\ref{fig:all} convolutional bidirectional NNs, specifically,
CBLSTM constitute the second most efficient
set of ANN models.
In contrast,
because of the longer training times and not much improved accuracy, bidirectional recurrent ANN models are the least efficient.

\section{\label{sec:conclusions}Conclusions}
We performed a benchmark study of 22 supervised machine learning methods comparing their ability
to accurately forecast long-time dynamics of a two-level quantum system linearly coupled to harmonic bath. We illustrate
that many of the studied ML methods can achieve a good agreement with the exact population
dynamics. Our study reveals that if the memory
time of a quantum dynamical system under study is known, the models based on Kernel Ridge Regression are the most accurate and
should be preferred. The commonly employed Gaussian kernel is confirmed to be the most efficient yielding 
the accuracy on a par with other nonlinear kernels while requiring somewhat less than training time.

Convolutional recurrent neural networks appear to be the most promising ANN models. 
One scenario where such models are suitable
are the problems when the memory time of the problem is not available and flexible ML models allowing variable size input are needed.
Such investigation is beyond the scope of this work and will be performed and reported in future studies. 
Based on the present study we conclude that particularly CGRU is the most promising ANN models for long-time quantum
dynamics simulations.

All ANN models are however less
accurate than KRR methods with nonlinear kernels and take more time to train and predict. 
Often quoted poor computational scaling of KRR with the increasing system size may pose some technical problems for large data sets which may be needed for large systems, longer and larger number of training trajectories. Nevertheless, many approaches have been suggested to mitigate this problem\cite{snelson2005sparse,deringer2021gaussian,rahimi2007random,yu2016orthogonal,hu2018inclusion,browning2022gpu} and we are also working on implementation of alternative approaches for large data sets.

One can reasonably anticipate that our results might depend on the strategy used to compare ML models based on
artificial neural networks. Here we chosen to build and compare ANN models with approximately the same number of
trainable parameters.
There is however no proven best way to compare the performance of ANNs with fundamentally different types of layers.
 We believe that the strategy chosen in this work should provide a faithful comparison of ANN models.
Additionally, our conclusion advocating KRR methods holds irrespective of the strategy used to compare ANN models
because KRR models studied in this work, are both faster and more accurate than ANN models. 
Addition of more layers to ANN models will likely
make them more accurate (although overfitting should be carefully checked in this case) but it will necessarily make
such models more computationally expensive.

FFNN may seem as a reasonable compromise between accuracy and computational cost.
However, one needs to bear in mind that
FFNN models require a fixed-size input. 
Of course an input to FFNN models can be padded with zeros, but it necessarily alters the representation of the
underlying physics in the input data and, therefore,
the performance of such models is not expected to be strong.

Many popular ML methods have been studied in the present article. However, the field of artificial intelligence and
machine learning is growing rapidly making it nearly impossible to cover all recently developed algorithms. Our future work
will focus on novel approaches to time-series modeling such as transformers.~\cite{vaswani17} 
Additionally, the CNN models employed in this work uses kernels of the same size. 
Even though the kernel size was optimized for the given task,
the use of inception modules~\cite{szegedy15} that include multiple filters of varying size might improve the performance of
CNNs and will also be tested. 

\begin{acknowledgments}
A.A.K. acknowledges the Ralph E. Powe Junior Faculty Enhancement Award from Oak Ridge Associated Universities. 
This work was also supported by the startup funds of the College of Arts and Sciences and
the Department of Physics and Astronomy of the University of Delaware. P.O.D. acknowledges funding by the National Natural Science Foundation of China (No. 22003051), the Fundamental Research Funds for the Central Universities (No. 20720210092) and via the Lab project of the State Key Laboratory of Physical Chemistry of Solid Surfaces.
K. J. R. E. acknowledges support by the Beyond Research Program between University of Delaware and Universidad Nacional de Colombia. Calculations were
performed with high-performance computing resources provided by the University of Delaware and Xiamen University.
\end{acknowledgments}

\section*{Author declarations}
\subsection*{Conflict of interest}
The authors have no conflicts to disclose.

\section*{Data availability}
The data that support the findings of this study are available
from the corresponding authors upon reasonable request.

\appendix
\section{\label{app:krrsubtrain}Comparison of performance of KRR models trained on sub-training and training sets}

As obvious from Table~\ref{tab:3}, KRR models trained only on the sub-training set have accuracy very close to the models trained on the entire training set (reported in the main text). It is desirable to include all points to ensure that no points are ''wasted''.

\begin{table*}
\caption{Mean absolute prediction errors (MAEs), training, and 
average single step prediction times of all KRR models used in this work. 
KRR-L, KRR-G, KRR-DP, KRR-E, and KRR-M ($n=1,2,3,4$) denote  kernel ridge regression models with linear kernel, 
Gaussian kernel, decaying-periodic kernel, exponential kernel, 
and Matern kernel with $n = 1,2,3,4$, respectively. The MAEs are shown for both symmetric and asymmetric cases. KRR models trained on 100\% training data and sub-training data (80\% of training data).}  
\begin{ruledtabular}
\begin{tabular}{l*{2}c} 
   & \multicolumn{2}{c}{Mean Absolute Prediction Error} \\ \cline{2-3}
Model   &  \multicolumn{1}{c}{Symmetric (100 $|$ 80)\%} & \multicolumn{1}{c}{Asymmetric (100 $|$ 80)\%} \\ \hline
KRR-L & (1.2 $|$ 1.2)$\cdot10^{-2}$  & (6.5 $|$ 5.8)$\cdot10^{-2}$ \\
 KRR-G  & (4.7 $|$ 3.8)$\cdot10^{-4}$  & (1.2 $|$ 1.3)$\cdot10^{-3}$ \\
 KRR-DP  & (4.3 $|$ 4.5)$\cdot10^{-4}$  & (2.0 $|$ 2.2)$\cdot10^{-3}$ \\
 KRR-E  & (2.1 $|$ 2.5)$\cdot10^{-3}$  & (3.3 $|$ 3.9)$\cdot10^{-3}$ \\
KRR-M1  & (2.4 $|$ 2.0)$\cdot10^{-4}$  & (1.3 $|$ 1.6)$\cdot10^{-3}$  \\
KRR-M2 & (2.2 $|$ 3.2)$\cdot10^{-4}$ & (2.7 $|$ 2.6)$\cdot10^{-3}$ \\
KRR-M3  & (2.0 $|$ 2.3)$\cdot10^{-4}$ & (2.3 $|$ 2.1)$\cdot10^{-3}$ \\
KRR-M4   & (2.3 $|$ 2.4)$\cdot10^{-4}$ & (2.1 $|$ 2.0)$\cdot10^{-3}$ 
\end{tabular}
\end{ruledtabular}
\label{tab:3}
\end{table*}

\section{\label{app:pso}Particle swarm optimization}
The hyperparameter optimization of the 1D CNN model was performed using the heuristic particle swarm optimization (PSO) algorithm.~\cite{kennedy95}
The PSO algorithm was inspired by the observation of the motion of swarms of birds and insects, where each swarm member is guided 
not only by the best solution for itself but also by the best solution seen by the entire population. 

An object of a swarm that moves around in the search space is called a particle. A new update of the swarm is called a new generation. The particle in the swarm includes variables (hyperparameters of the model) updated during the optimization based on
the information about the previous best (test MAE) states of the particle and the swarm itself. Each variable of an individual particle has velocity $\mathbf{v}(t)$, position $\mathbf{x}(t)$, and the best particle's position  $\mathbf{x}_p(t)$
which has generated the smallest MAE during the coarse of its trajectory. Additionally, there is a global variable that contains the best global position $\mathbf{x}_g(t)$ of the swarm. 
Each generation the position and velocity of a particle are updated according to 
\begin{equation}
\begin{array}{l}
\mathbf{v}_{i}(t+1)=w \mathbf{v}_{i}(t)+c_{p}\left(\mathbf{x}_{p_{i}}(t)-\mathbf{x}_{i}(t)\right)+c_{g}\left(\mathbf{x}_{g_{i}}(t)-\mathbf{x}_{i}(t)\right) \\
\mathbf{x}_{i}(t+1)=\mathbf{x}_{i}(t)+\mathbf{v}_{i}(t),
\end{array}
\label{eq:PSO}
\end{equation}
where $c_p$, $c_g$, and $w$ are cognitive, social, and inertia coefficients. These coefficients quantify how much the particle is directed toward the best solution seen by itself, by the swarm, and in the previous direction.

In the present work, the search space is conformed by the space of hyperparameters of the 1D CNN model as described
in Sec.~\ref{sec:opt}. 
The values of the parameters are restricted to be positive at each step. If a parameter takes a negative value it is replaced by a random number in an interval of values initially set for the parameter. 
The following values of the coefficients are used $w = 0.729$, $c_p = 1.49445$, $c_g = 1.49445$.

\bibliography{refs}

\end{document}